\documentclass[aps,pre,twocolumn,10pt,superscriptaddress,nofootinbib,balancelastpage,dvipsnames]{revtex4-2} 
\usepackage[latin1]{inputenc}
\usepackage{amsmath,amssymb}
\usepackage{mathrsfs} 
\usepackage{hyperref}
\usepackage[capitalise]{cleveref}
\usepackage{siunitx}
\usepackage{braket}
\usepackage{calc}
\usepackage{esint}
\usepackage{tabularx}
\usepackage{dsfont}
\usepackage{color}
\usepackage{ifthen}
\usepackage{graphicx}
\usepackage{seqsplit}
\allowdisplaybreaks

\usepackage{tikz}

\newcommand{\orcidicon}{\includegraphics[width=0.32cm]{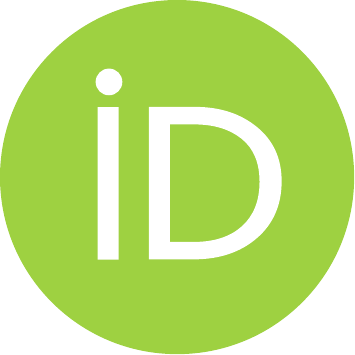}}

\foreach \x in {A, ..., Z}{%
\expandafter\xdef\csname orcid\x\endcsname{\noexpand\href{https://orcid.org/\csname orcidauthor\x\endcsname}{\noexpand\orcidicon}}
}

\newcommand{\Eins}{\mathds{1}}%

\newcommand{\Kronecker}[2]{\delta_{#1#2}}%
\newcommand{\ii}{\mathrm{i}}%

\newcommand{\dif}{\mathrm{d}}%
\newcommand{\Nabla}{\vec{\nabla}}%
\newcommand{\fdif}{\operatorname{\delta}}%
\newcommand{\Fdif}[2]{\frac{\fdif\!#1}{\fdif\!#2}}%

\newcommand{\uu}{\hat{u}}

\newcommand{\norm}[1]{\lVert#1\rVert}%

\newcommand{\ZT}[1]{\textquotedblleft#1\textquotedblright}%
\newcommand{\Tensor}[1]{\underline{\boldsymbol{#1}}}

\newcolumntype{Y}{>{\centering\arraybackslash}X}%
\newcolumntype{Z}{>{\raggedright\arraybackslash}X}%

\newlength{\myl}%
\newcommand{\SUM}[2]{{\setlength{\myl}{\widthof{$\displaystyle\sum_{#1}^{#2}$}*\real{0.5}-\widthof{$\displaystyle\sum$}*\real{0.5}}\sum_{#1}^{#2}\;\hspace{-\the\myl}}}
\newcommand{\INT}[3]{\settowidth{\myl}{$\displaystyle\int_{#1}^{#2}$}{\int_{#1}^{#2}\;\;\;\hspace{-\the\myl}\dif #3}\,}
\newcommand{\TINT}[3]{\settowidth{\myl}{$\int_{#1}^{#2}$}{\int_{#1}^{#2}\!\ifthenelse{\equal{#1#2}{}}{}{\;\;\;\;\hspace{-\the\myl}}\dif #3}\,}%
\newcommand{\EINT}[3]{\settowidth{\myl}{$\int_{#1}^{#2}$}{\int_{#1}^{#2}\;\;\;\,\hspace{-\the\myl}\dif #3}\,}
\newcommand{\CINT}[3]{\settowidth{\myl}{$\displaystyle\int_{#1}^{#2}$}{\oint_{#1}^{#2}\;\;\;\hspace{-\the\myl}\dif #3}\,}

\newcommand{\ERAI}{\mathrel{\phantom{=}}\negmedspace{}}%
\newcommand{\ppartial}[1]{{\partial^{2}_{#1}}}%
\DeclareMathOperator\arctanh{arctanh}%

\begin{document}
\title{From a microscopic inertial active matter model to the Schr\"odinger equation}

\author{Michael te Vrugt\orcidA{}}
\affiliation{Institut f\"ur Theoretische Physik, Westf\"alische Wilhelms-Universit\"at M\"unster, 48149 M\"unster, Germany}
\affiliation{Center for Soft Nanoscience (SoN), Westf\"alische Wilhelms-Universit\"at M\"unster, 48149 M\"unster, Germany}

\author{Tobias Frohoff-H\"ulsmann\orcidB{}}
\affiliation{Institut f\"ur Theoretische Physik, Westf\"alische Wilhelms-Universit\"at M\"unster, 48149 M\"unster, Germany}

\author{Eyal Heifetz\orcidC{}}
\affiliation{Porter School of the Environment and Earth Sciences, Tel Aviv University, 69978 Tel Aviv, Israel}

\author{Uwe Thiele\orcidD{}}
\email{u.thiele@uni-muenster.de}
\homepage{http://www.uwethiele.de}
\affiliation{Institut f\"ur Theoretische Physik, Westf\"alische Wilhelms-Universit\"at M\"unster, 48149 M\"unster, Germany}
\affiliation{Center for Nonlinear Science (CeNoS), Westf\"alische Wilhelms-Universit\"at M\"unster, 48149 M\"unster, Germany}
\affiliation{Center for Multiscale Theory and Computation (CMTC), Westf\"alische Wilhelms-Universit\"at M\"unster, 48149 M\"unster, Germany}

\author{Raphael Wittkowski\orcidE{}}
\email[Corresponding author: ]{raphael.wittkowski@uni-muenster.de}
\affiliation{Institut f\"ur Theoretische Physik, Westf\"alische Wilhelms-Universit\"at M\"unster, 48149 M\"unster, Germany}
\affiliation{Center for Soft Nanoscience (SoN), Westf\"alische Wilhelms-Universit\"at M\"unster, 48149 M\"unster, Germany}
\affiliation{Center for Nonlinear Science (CeNoS), Westf\"alische Wilhelms-Universit\"at M\"unster, 48149 M\"unster, Germany}

\begin{abstract}
Field theories for the one-body density of an active fluid, such as the paradigmatic active model B+, are simple yet very powerful tools for describing phenomena such as motility-induced phase separation. No comparable theory has been derived yet for the underdamped case. In this work, we introduce \textit{active model I+}, an extension of active model B+ to particles with inertia. The governing equations of active model I+ are systematically derived from the microscopic Langevin equations. We show that, for underdamped active particles, thermodynamic and mechanical definitions of the velocity field no longer coincide and that the density-dependent swimming speed plays the role of an effective viscosity. Moreover, active model I+ contains the Schr\"odinger equation in Madelung form as a limiting case, allowing to find analoga of the quantum-mechanical tunnel effect and of fuzzy dark matter in the active fluid. We investigate the active tunnel effect analytically and via numerical continuation.
\end{abstract}
\maketitle

\section{Introduction}
The study of active particles has become one of the fastest-growing fields of research in soft matter physics and statistical mechanics due the enormous number of novel and interesting effects that active matter can exhibit. Among these effects are a plethora of analogies between active matter and quantum mechanics. This includes Bose-Einstein condensation \cite{MengMMG2021,Golestanian2019,MahaultG2020}, Hall viscosities \cite{HanFSVIdPV2020,MarkovichL2021}, orientational order in systems of fully symmetric particles \cite{teVrugtW2020b,teVrugtW2019c}, Schr\"odinger-type dynamics in polar liquids \cite{SouslovvZBV2017}, spin-orbit coupling \cite{LoeweSG2018}, time crystals \cite{EversW2021}, and topological effects \cite{PalaciosTGPSG2021}. A very simple yet extremely powerful description for active matter is given by scalar active field theories such as \textit{active model B} (AMB) \cite{WittkowskiTSAMC2014} and the more general \textit{active model B+} (AMB+) \cite{TjhungNC2018}. These provide a minimal description for many remarkable effects such as active phase separation and have led to crucial insights into the thermodynamics of active matter \cite{MarkovichFTC2021,CaballeroC2020,NardiniFTvWTC2017,SinghC2019,Cates2019}.
 
While such field theories have also been coupled to the momentum-conserving dynamics of the solvent \cite{TiribocchiWMC2015,NardiniFTvWTC2017,SinghC2019,KoleARM2021}, the inertia of the active particles \textit{themselves} has been ignored in this context. However, recent experiments \cite{ScholzJLL2018,LeoniPEENASA2020,TapiaGS2021} have found that the inertia of active particles is important in a variety of contexts. Moreover, theoretical and experimental studies have found a number of remarkable effects associated with inertial active matter \cite{Loewen2020}, ranging from self-sustained temperature gradients \cite{MandalLL2019} through restored equilibrium crystallization \cite{OmarKGGB2021} to damping-dependent phase boundaries \cite{teVrugtJW2021}. Consequently, there has been a strongly increasing recent interest in inertial active matter  \cite{CapriniM2021,Sandoval2020,SprengerJIL2021,NguyenWL2021,SuJH2021}.
 
Field theories for inertial active matter have been derived in Refs.\ \cite{teVrugtJW2021,AroldS2020,AroldS2020b} as extensions of the active phase field crystal (PFC) model \cite{MenzelL2013,MenzelOL2014,OphausGT2018,OphausKGT2021}. Active PFC models can be derived as an approximation of dynamical density functional theory (DDFT) \cite{teVrugtLW2020}, and have two disadvantages compared to AMB+. First, they rely on the close-to-equilibrium (\ZT{adiabatic}) approximation that DDFT is based on, and second, they require two order parameter fields (density $\rho$ and polarization $\vec{P}$) rather than just one, making them more complex. In contrast to PFC models, to the best of our knowledge, no extension of AMB+ to the underdamped case has been derived yet. A second gap is that up to now the relation between \textit{inertial} active matter and quantum mechanics remains unexplored. 
 
In this work, we close both of these gaps. First, we obtain, via a microscopic derivation using the well-established interaction expansion method \cite{WittkowskiSC2017,BickmannBJW2020,BickmannW2019b,BickmannW2020}, an extension of AMB+ to particles with inertia that we refer to as \textit{active model I+}. The derivation systematically generalizes previous work on this subject and also reveals some remarkable new physics. In particular, it is found that thermodynamic and mechanical definitions of the velocity field lead to different results in the active case, and that the density-dependent swimming speed of active particles gives rise to an effective viscosity of the active fluid. Second, we show that active model I (the underdamped analogon of AMB) contains, as a limiting case, the Madelung equations \cite{Madelung1926,Madelung1927}, which constitute a hydrodynamic representation of the Schr\"odinger equation \cite{HeifetzC2015,TsekovHC2017}. This allows us to find analoga of the quantum-mechanical tunnel effect and of fuzzy dark matter in an active fluid. A numerical investigation of the active tunnel effect using continuation methods shows that it also occurs when the approximations required for the active-quantum mapping are not exactly satisfied. This implies its robustness as physical phenomenon.

\section{Results}
\subsection{\label{amiplus}Active model I+}
Our starting point is AMB+ \cite{TjhungNC2018}, which is given by
\begin{equation}
\dot{\rho} = \Nabla\cdot \big(M\rho (\Nabla(f'_o(\rho) - \kappa \Nabla^2 \rho + \lambda(\Nabla \rho)^2) - \xi(\Nabla^2\rho)\Nabla\rho) \big)
\label{activemodelb}
\end{equation}
with a local particle number density $\rho(\vec{r},t)$, a mobility $M$, an (overdamped) free-energy density $f_o(\rho)$ typically assumed to be a fourth-order polynomial, the notation $f'_o = \partial_\rho f_o$, and constants $\kappa$, $\lambda$, and $\xi$. An overdot denotes a partial derivative with respect to time $t$. The model \eqref{activemodelb} is overdamped. Typically, one makes the simplifying assumption of a constant mobility $M\rho \approx M_0$, which is valid only in uniform states or close to a critical point, but qualitatively reasonable also in other cases. The purpose of this approximation is to get a noise that is additive rather than multiplicative \cite{Cates2019}. 
Here, we do not have a noise term since our microscopic derivation interprets $\rho$ as an ensemble-averaged density \cite{ArcherR2004}. By setting $\xi=0$ in \cref{activemodelb}, one obtains AMB \cite{WittkowskiTSAMC2014}. AMB, in turn, can be thought of as a minimal extension of the Cahn-Hilliard equation \cite{Cahn1965} to the active case.
 
A remarkable feature of AMB and AMB+, which distinguishes them from passive field theories, is that the right-hand side of \cref{activemodelb} cannot be written as a gradient dynamics, i.e., in terms of the functional derivative of a free energy \cite{WittkowskiTSAMC2014,StenhammarTAMC2013}. In addition, AMB+ allows (unlike AMB) for circulating currents in steady state \cite{TjhungNC2018}. One can derive AMB+ either phenomenologically by writing down a general theory of a certain order in gradients and fields (top-down approach) or microscopically by explicit coarse-graining of the microscopic equations of motion of active particles (bottom-up approach) \cite{Cates2019}. Here, the bottom-up approach has the advantage of providing explicit expressions for the coefficients appearing in the model (predictive theory) \cite{BickmannW2020,BickmannW2019b} and giving a clearer insight into the origin of the various terms and the approximations required to get them.
 
Since AMB+ is overdamped, it does not take the inertia of the active particles into account. In this work, we obtain via a microscopic derivation an extension of AMB+ to the underdamped case, which we will refer to as \textit{active model I+} (AMI+), with \ZT{I} standing for \ZT{inertial}. It is given by
\begin{gather}
\dot{\rho} = -\Nabla \cdot (\rho \vec{v}) + \frac{1}{2 D_R}\Nabla\cdot(v_{\mathrm{ld}}(\rho)\Nabla v_{\mathrm{ld}}(\rho)\rho),\label{dotrhogeneral}\\
\begin{split}
\dot{\vec{v}} + (\vec{v}\cdot\Nabla)\vec{v} &= -\frac{1}{m}\Nabla(f'(\rho)- \kappa \Nabla^2 \rho + \lambda(\Nabla \rho)^2+ U_1) \\&\ERAI{}- \gamma \vec{v} + \frac{v_{\mathrm{ld}}(\rho)^2}{\gamma}\Nabla^2\vec{v} +\xi(\Nabla^2\rho)\Nabla\rho
\end{split}\raisetag{3em}\label{dotvgeneral}
\end{gather}
with the velocity field $\vec{v}$, the rotational diffusion coefficient $D_R$, the free energy density $f$, its derivative $f'=\partial_\rho f$, the particle mass $m$, the friction coefficient $\gamma = 1/(mM)$, and the local density-dependent swimming speed
\begin{equation}
v_{\mathrm{ld}}(\rho)=v_0 -\frac{A_1}{\gamma m}\rho.
\label{vdrho}
\end{equation}
Here, $v_0$ is the propulsion speed of a free particle and $A_1$ is a constant (see \cref{a1} in \cref{mikro}). We have also included an external potential $U_1$ for generality. The form \eqref{vdrho} agrees with the expression derived in Ref.\ \cite{BickmannW2020}. In overdamped fluids, the existence of a density-dependent swimming speed -- that can arise, e.g., from particle collisions in the case of active Brownian particles (ABPs) considered here or from quorum-sensing in the case of bacteria -- is essential for the phenomenon of \textit{motility-induced phase separation} (MIPS), where repulsively interacting particles phase-separate (which would not be possible in a passive system) \cite{CatesT2015}. 
From AMI+, we can see that $v_{\mathrm{ld}}(\rho)$ plays two interesting roles in the underdamped case:
\begin{enumerate}
    \item It leads to a second term in the continuity equation \eqref{dotrhogeneral} in addition to the well-known passive term $\Nabla \cdot (\rho \vec{v})$. This second term is related to the self-propulsion term known from the active PFC model (see \cref{qsa}).
    \item It gives rise to an effective viscosity\footnote{\ZT{Effective viscosities} have been discussed in active matter also in other contexts, see Ref.\ \cite{BaerGHP2020} for a recent overview.} $v_{\mathrm{ld}}(\rho)^2/\gamma$. This implies that a system of underdamped active particles should behave more like a viscous fluid for larger activity (larger $v_{\mathrm{ld}}$) and more like an ideal fluid for larger density (smaller $v_{\mathrm{ld}}$). 
\end{enumerate}
AMI+ contains AMB+ as a limiting case. Showing this requires two approximations:
\begin{enumerate}
\item We assume the system to be overdamped (large $\gamma$), i.e., we set the material derivative $\dot{\vec{v}} + (\vec{v}\cdot\Nabla)\vec{v}$ in \cref{dotvgeneral} to zero, solve the resulting equation for $\vec{v}$ and insert the result into \cref{dotrhogeneral}. (This is analogous to the procedure required for deriving overdamped from underdamped DDFT \cite{teVrugtLW2020}.)
\item Using \cref{vdrho}, we write in \cref{dotrhogeneral}
\begin{equation}
\frac{1}{2 D_R}\Nabla\cdot(v_{\mathrm{ld}}(\rho)\Nabla v_{\mathrm{ld}}(\rho)\rho) = \Nabla\cdot(M\rho \Nabla f'_e(\rho)) + \mathcal{O}(\rho^3)
\end{equation}
with the effective free energy density
\begin{equation}
f_e = \frac{1}{2MD_R}\bigg(v_0^2\rho\big(\ln(\Lambda^2\rho)-1\big)-\frac{3v_0 A_1}{2\gamma m}\rho^2\bigg),\label{effectivefree}
\end{equation}
where $\Lambda$ is the (irrelevant) thermal de Broglie wavelength, and then define $f_o = f + f_e$ and $f'_e = \partial_\rho f_e$. Equation \eqref{effectivefree} shows that we can interpret $v_0^2/(2MD_R)$ as a shift of the temperature \cite{PreislerD2016}, since the first term on the right-hand side has the form of an ideal gas free energy.
\end{enumerate}
Interestingly, as shown in \cref{qsa}, the microscopic derivation reveals another form of \ZT{effective temperature} that is a feature of inertial active matter. The free energy $f$ appearing in \cref{dotvgeneral} and microscopically given by \cref{fmicro} has, as a prefactor in the ideal gas term, a factor $k_BT + mv_0^2/2$ rather than $k_B T$ as in the passive case. This shows that the active kinetic energy $mv_0^2/2$ plays the role of a thermal energy in inertial active matter.

By taking the curl of \cref{dotvgeneral} and defining the vorticity $\vec{\omega} = \Nabla\times\vec{v}$, we can obtain the active vorticity equation
\begin{equation}
\begin{split}
\dot{\vec{\omega}}  &=-(\vec{v}\cdot\Nabla)\vec{\omega} +(\vec{\omega}\cdot\Nabla)\vec{v}-\vec{\omega}(\Nabla\cdot\vec{v})+\frac{v_{\mathrm{ld}}^2(\rho)}{\gamma}\Nabla^2\vec{\omega} 
\\&\ERAI{}+ \frac{1}{\gamma}(\Nabla v^2_{\mathrm{ld}}(\rho))\times\Nabla^2\vec{v} - \gamma\vec{\omega} -\xi(\Nabla\rho)\times(\Nabla\Nabla^2\rho).     
\end{split}
\end{equation}

Starting from AMI+, we can again make two approximations:
\begin{enumerate}
    \item We assume that $v_{\mathrm{ld}}(\rho)$ is small. From \cref{vdrho}, we can see that this assumption is justified if $v_0$ and $A_1$ are both small (i.e., in the case of weak activity) or, for larger activities, if $v_0\approx A_1\rho/(\gamma m)$.
    \item We drop the term proportional to $\xi$, such that the material derivative of $\vec{v}$ is given by the sum of the gradient of a generalized chemical potential and a damping term. Setting $\xi=0$ is the usual approximation by which one gets from AMB+ to AMB.
\end{enumerate}
We then obtain the simpler \textit{active model I} (AMI), which is given by
\begin{align}
\dot{\rho} &= -\Nabla \cdot (\rho \vec{v}),\label{activemodelbu1}\\
\dot{\vec{v}} + (\vec{v}\cdot\Nabla) \vec{v} &= -\frac{1}{m}\Nabla(f'(\rho)- \kappa \Nabla^2 \rho + \lambda(\Nabla \rho)^2 + U_1) - \gamma \vec{v}.
\label{activemodelbu2}
\end{align}
Equation \eqref{activemodelbu2} can be written as 
\begin{equation}
\dot{\vec{v}} + (\vec{v}\cdot\Nabla) \vec{v} = -\frac{1}{m}\Nabla\mu - \gamma \vec{v}    
\end{equation}
with a generalized chemical potential
\begin{equation}
\mu = f'(\rho) - \kappa \Nabla^2\rho +\lambda (\Nabla\rho)^2 + U_1.  
\label{static}
\end{equation}
It is straightforward to obtain AMB from AMI by taking the overdamped limit.

\subsection{\label{mikro}Microscopic derivation of active model I+}
Microscopically, a two-dimensional system of $N$ underdamped ABPs is described by the Langevin equations \cite{AroldS2020}
\begin{align}
\dot{\vec{r}}_i &= \frac{\vec{p}_i}{m},\label{langevin1}\\    
\dot{\vec{p}}_i &= - \gamma \vec{p}_i - \Nabla_{\vec{r}_i} U(\{\vec{r}_i\}) + m\gamma v_0 \uu_i + \sqrt{2D}\vec{\eta}_i\label{langevin2},\\
\dot{\varphi}_i &= \sqrt{2D_R}\xi_i\label{langevin3},
\end{align}
where $\vec{r}_i(t)$, $\vec{p}_i(t)$, and $\varphi_i(t)$ are position, momentum, and orientation (direction of self-propulsion force) of the $i$-th particle, $\uu(\varphi) = (\cos(\varphi),\sin(\varphi))^\mathrm{T}$ is its orientation vector, $m$ is its mass, $\gamma$ and $\gamma_R$ are the translational and rotational friction coefficients, $D$ and $D_R$ are the translational and rotational diffusion coefficients, $U = U_2 + U_1$ the potential consisting of interaction potential $U_2$ and external potential $U_1$, $\vec{\eta}_i(t)$ and $\xi_i(t)$ are translational and rotational Gaussian white noises with zero mean and unit variance, and $v_0$ is the particles' self-propulsion velocity. The corresponding Fokker-Planck equation is given by \cite{AroldS2020}
\begin{equation}
\partial_t P_N(\{\vec{r}_i,\vec{p}_i,\varphi_i\}) = \ii L (\{\vec{r}_i,\vec{p}_i,\varphi_i\})P_N(\{\vec{r}_i,\vec{p}_i,\varphi_i\}),\label{fokkerplanck}     
\end{equation}
where $P_N$ is the $N$-body probability distribution and 
\begin{equation}
\begin{split}
&\ii L (\{\vec{r}_i,\vec{p}_i,\varphi_i\})
\\&=\sum_{i=1}^{N}\bigg(-\frac{\vec{p}_i}{m}\cdot \Nabla_{\vec{r}_i} + \gamma+\gamma\vec{p}_i\cdot\Nabla_{\vec{p}_i}+ (\Nabla_{\vec{r}_i}U)\cdot \Nabla_{\vec{p}_i} \\
&\ERAI{} - m\gamma v_0 \uu_i\cdot\Nabla_{\vec{p}_i}+ D\Nabla_{\vec{p}_i}^2 + D_R \ppartial{\varphi_i}\bigg)
\end{split}
\end{equation}
is the Liouvillian. The dependence on $t$ is not written explicitly to simplify the notation.

By integrating \cref{fokkerplanck} over the coordinates of all except for one particle, we find \cite{AroldS2020}
\begin{align}
&\partial_t P_1(\vec{r},\vec{p},\uu)
\nonumber\\&=\bigg(-\frac{\vec{p}}{m}\cdot\Nabla + \gamma + \gamma\vec{p}\cdot\Nabla_{\vec{p}}+(\Nabla U_1) \cdot\Nabla_{\vec{p}}\label{f1}
\nonumber\\&\ERAI{} - m\gamma v_0 \uu\cdot\Nabla_{\vec{p}} + D \Nabla_{\vec{p}}^2 + D_R\partial_\varphi^2\bigg)P_1(\vec{r},\vec{p},\uu) 
\\&\ERAI{}+\INT{}{}{^2r_2}\INT{}{}{^2p_2}\INT{}{}{\varphi_2}(\Nabla U_2)\cdot\Nabla_{\vec{p}}P_2(\vec{r},\vec{r}_2,\vec{p},\vec{p}_2,\varphi,\varphi_2) 
\nonumber
\end{align}
with $\Nabla = \Nabla_{\vec{r}}$ and the one- and two-particle distribution functions $P_1$ and $P_2$ defined as \cite{Archer2009}
\begin{align}
\nonumber P_{n}&=\frac{N!}{(N-n)!}\INT{}{}{^2 r_1}\dotsb\INT{}{}{^2 r_{N-n}}\INT{}{}{^2 p_1}\dotsb\INT{}{}{^2 p_{N-n}}
\\&\ERAI{}\INT{}{}{\varphi_1}\dotsb\INT{}{}{\varphi_{N-n}}P_N,    
\end{align}
where $n\in\{1,\dotsc,N\}$. The index 1 is dropped for the coordinates. We define the particle density
\begin{equation}
\varrho(\vec{r},\uu)=\INT{}{}{^2 p}P_1(\vec{r},\vec{p},\uu) 
\label{density}
\end{equation}
and apply the local equilibrium approximation \cite{Archer2009}
\begin{equation}
P_1(\vec{r},\vec{p},\uu)=\frac{\varrho(\vec{r},\uu)}{2\pi m k_B T}\exp\!\bigg(-\frac{(\vec{p}-m\mathfrak{\vec{v}})^2}{2mk_B T}\bigg)
\label{localeq}
\end{equation}
with the Boltzmann constant $k_B$, the temperature $T$, and a generalized velocity field $\mathfrak{\vec{v}}(\vec{r},\uu)$. 
The expression \eqref{localeq} may also be seen as the zeroth-order term of an expansion of a general distribution $P_1$ around local equilibrium; the higher-order terms then give viscous corrections \cite{Archer2009}. Equation \eqref{localeq} implies
\begin{equation}
\varrho(\vec{r},\uu)\mathfrak{\vec{v}}(\vec{r},\uu)=\INT{}{}{^2p}\frac{\vec{p}}{m}P_1(\vec{r},\vec{p},\uu).
\label{current}
\end{equation}
Although \cref{localeq} is a common approximation, the fact that we allow $\mathfrak{\vec{v}}$ to depend on $\uu$ (i.e., that we use a generalized velocity) distinguishes our approach from previous passive \cite{Archer2009} and active \cite{AroldS2020} inertial field theories.

We now drop arguments of the fields unless unclear. Integrating \cref{f1} over $\vec{p}$ and using \cref{density,current} yields
\begin{equation}
\partial_t\varrho=-\Nabla\cdot(\varrho\mathfrak{\vec{v}}) + D_R \partial_\varphi^2  \varrho\label{dotvarrho}.  
\end{equation}
Similarly, we can multiply \cref{f1}  by $\vec{p}/m$, integrate over $\vec{p}$, and use \cref{density,localeq,current,dotvarrho} to get
\begin{align}
\partial_t\mathfrak{\vec{v}}+(\mathfrak{\vec{v}}\cdot\Nabla)\mathfrak{\vec{v}}&= -D_R\mathfrak{\vec{v}}\frac{\partial_\varphi^2 \varrho}{\varrho}- \gamma \mathfrak{\vec{v}} + \gamma v_0\uu - \frac{1}{m}\Nabla U_1
\nonumber\\&\ERAI{} -\frac{k_B T}{m}\Nabla\ln(\Lambda^2\varrho)- \frac{1}{m\varrho}\vec{\mathcal{I}}  
\label{partialu}
\end{align}
with the interaction term
\begin{equation}
\vec{\mathcal{I}}=\INT{}{}{^2r_2}\INT{}{}{\varphi_2}\varrho_2(\vec{r},\vec{r}_2,\varphi,\varphi_2)\Nabla U_2(r),
\label{interactionpart}
\end{equation}
where $\varrho_2 = \TINT{}{}{^2p}\TINT{}{}{^2p_2}P_2$ is the two-body distribution function and $r=\norm{\vec{r}-\vec{r}_2}$ with the Euclidean norm $\norm{\cdot}$ is a distance. The derivation of \cref{partialu} generally follows the standard procedure of deriving hydrodynamic equations from microscopic dynamics \cite{Archer2009}. Our result differs from the standard form of velocity transport equations by the presence of the term $-D_R\mathfrak{\vec{v}}(\partial_\varphi^2 \varrho)/\varrho$, which arises from the term $D_R \partial_\varphi^2\varrho$ in \cref{dotvarrho}.

We can define the pair-distribution function $g$ as \cite{BickmannW2020,WittkowskiSC2017}
\begin{equation}
g(\vec{r},\vec{r}_2,\varphi,\varphi_2)=\frac{\varrho_2(\vec{r},\vec{r}_2,\varphi,\varphi_2)}{\varrho(\vec{r},\varphi)\varrho(\vec{r}_2,\varphi_2)}.
\end{equation}
Following the treatment in Ref.\ \cite{BickmannW2020}, we assume the pair-distribution function to be translationally and rotationally invariant, implying that it can be written as $g(r,\theta_1,\theta_2)$ with the angles $\theta_1 = \varphi_R-\varphi$ and $\theta_2 = \varphi_2 -\varphi$ and the parametrization $\vec{r}-\vec{r}_2 = r\uu(\varphi_R)$. Then, we can perform a Fourier and a gradient expansion \cite{YangFG1976,BickmannW2020} of $g$ and find
\begin{equation}
\begin{split}
\vec{\mathcal{I}}&=\sum_{l=0}^{\infty}\frac{1}{l!}\varrho(\vec{r},\phi,t)\INT{0}{\infty}{r}r^{l+1}U_2'(r)\INT{0}{2\pi}{\varphi_R}\uu(\varphi_R)(\uu(\varphi_R)\cdot\Nabla)^l 
\\&\ERAI{}\INT{0}{2\pi}{\varphi_2}\sum_{n_1,n_2 = -\infty}^{\infty} 
g_{n_1 n_2}(r)\cos(n_1\theta_1+n_2\theta_2)\varrho(\vec{r},\varphi_2,t)    
\end{split}\raisetag{1em}
\end{equation}
with the $r$-dependent expansion coefficients \cite{BickmannW2020}
\begin{equation}
g_{n_1 n_2}(r) = \frac{\TINT{0}{2\pi}{\theta_1}\TINT{0}{2\pi}{\theta_2}\, g(r, \theta_1, \theta_2)\cos(n_1\theta_1+n_2\theta_2)}{\pi^2(1+\Kronecker{n_1}{0})(1+\Kronecker{n_2}{0})}
\end{equation}
and $U_2'(r)=\dif U_2/\dif r$.

We now carry out the Cartesian orientational expansions \cite{teVrugtW2020b}
\begin{align}
\varrho(\vec{r},\uu)&=\rho(\vec{r})+\uu\cdot\vec{P}(\vec{r}),\label{expansion1}\\
\mathfrak{\vec{v}}(\vec{r},\uu)&=\vec{v}(\vec{r})+\uu\cdot \Tensor{v}_{\vec{P}}(\vec{r})\label{expansion2}
\end{align}
with the non-orientational particle density 
\begin{equation}
\rho(\vec{r})=\frac{1}{2\pi}\INT{0}{2\pi}{\varphi}\varrho(\vec{r},\uu),   
\end{equation}
the local velocity
\begin{equation}
\vec{v}(\vec{r})=\frac{1}{2\pi}\INT{0}{2\pi}{\varphi}\mathfrak{\vec{v}}(\vec{r},\uu),    
\end{equation}
the local polarization
\begin{equation}
\vec{P}(\vec{r})=\frac{1}{\pi}\INT{0}{2\pi}{\varphi}\uu\varrho(\vec{r},\uu),
\end{equation}
and the local velocity polarization 
\begin{equation}
 \Tensor{v}_{\vec{P}}(\vec{r})=\frac{1}{\pi}\INT{0}{2\pi}{\varphi}\uu\otimes\mathfrak{\vec{v}}(\vec{r},\uu)    
\end{equation}
with the dyadic product $\otimes$. Here, our treatment differs in an important way from standard treatments of active overdamped \cite{BickmannW2019b,BickmannW2020}, passive underdamped \cite{Archer2009} and even active underdamped \cite{AroldS2020} particles. Since we have a velocity field $\mathfrak{\vec{v}}$ that also depends on $\uu$, we have to perform the orientational expansion not only for the density, but also for the velocity.

We now insert \cref{expansion1} into $\ln(\Lambda^2\varrho)$ and Taylor expand around $\vec{P}=\vec{0}$. This gives
\begin{equation}
\ln(\Lambda^2\varrho) \approx \ln(\Lambda^2\rho) + \frac{1}{\rho}\uu\cdot\vec{P} \approx \ln(\Lambda^2\rho) +\frac{1}{\varrho_0}\uu\cdot\vec{P},\label{logtaylor}   
\end{equation}
where we have replaced $\rho$ by a spatially and temporally constant reference density $\varrho_0$ in the last step. Similarly, we insert \cref{expansion1,expansion2} into $\vec{\mathfrak{v}}(\partial_\varphi^2\varrho)/\varrho$ and Taylor expand around $\vec{P}=\vec{0}$ to find
\begin{equation}
\frac{(\vec{v}+\uu\cdot \Tensor{v}_{\vec{P}})\partial_\varphi^2(\rho+\uu\cdot\vec{P})}{\rho+\uu\cdot\vec{P}}\approx - \frac{(\vec{v}+\uu\cdot \Tensor{v}_{\vec{P}})\uu\cdot\vec{P}}{\rho}.\label{taylor3}
\end{equation}
Finally, an orientational expansion of the interaction term gives
\begin{equation}
\begin{split}
\frac{1}{\varrho}\vec{\mathcal{I}}&=A_1\uu\rho +A_2 \Nabla\rho + A_3\Nabla^2\vec{P}+2A_3\Nabla (\Nabla\cdot\vec{P})\\
&\ERAI{}+ A_4\uu\Nabla^2\rho + 2A_4 \Nabla (\Nabla\cdot\uu)\rho+ \dotsb    
\end{split}
\label{expandedinteraction}
\end{equation}
with the coefficients
\begin{align}
A_1 &= 2 \pi^2 \INT{0}{\infty}{r}r U_2'(r) (g_{1,0}(r) + g_{-1,0}(r)),\label{a1}\\
A_2 &= 2 \pi^2 \INT{0}{\infty}{r}r^{2}U_2'(r) g_{0,0}(r),\\
A_3 &= \frac{\pi^2}{4} \INT{0}{\infty}{r}r^3 U_2'(r) (g_{1,-1}(r) + g_{-1,1}(r)),\\
A_4 &= \frac{\pi^2}{2}  \INT{0}{\infty}{r}r^3 U_2'(r) (g_{1,0}(r) + g_{-1,0}(r)).
\end{align}
These coefficients can be time-dependent by inheriting a time-dependence of $g$ \cite{BickmannW2020}, but we will assume them to be constant.

From \cref{dotvarrho,partialu,expansion1,expansion2,expandedinteraction,logtaylor,taylor3}, we obtain the general \textit{local field theory for underdamped ABPs}
\begin{align}
\dot{\rho}&= - \Nabla \cdot(\rho\vec{v}) - \frac{1}{2}\Nabla\cdot(\vec{P}\cdot \Tensor{v}_{\vec{P}}),\label{generalconti}\\
\dot{\vec{P}}&= - \Nabla\cdot(\vec{v}\otimes\vec{P}) - \Nabla\cdot(\rho \Tensor{v}_{\vec{P}}) - D_R \vec{P},\label{dotP}\\
\dot{\vec{v}}&= - (\vec{v}\cdot\Nabla)\vec{v} - \frac{1}{2}( \Tensor{v}_{\vec{P}}\cdot\Nabla)\cdot \Tensor{v}_{\vec{P}} -\gamma \vec{v}-\frac{A_3}{m}\Nabla^2\vec{P}  \notag 
\\&\ERAI{}+D_R \frac{\vec{P}\cdot\Tensor{v}_{\vec{P}}}{2\rho} -\frac{1}{m}\Nabla \big(k_B T\ln(\Lambda^2\rho) + A_2 \rho \label{dotv}
\\&\ERAI{} + 2A_3(\Nabla\cdot\vec{P}) +U_1\big),\notag\\
\dot{\Tensor{v}}_{\vec{P}}&= -( \Tensor{v}_{\vec{P}}\cdot\Nabla)\otimes\vec{v} -(\vec{v}\cdot\Nabla) \Tensor{v}_{\vec{P}}-\gamma \Tensor{v}_{\vec{P}}+ \gamma v_0\Eins 
\notag\\&\ERAI{} +D_R \frac{\vec{v}\otimes\vec{P}}{\rho} -\frac{k_B T}{\varrho_0 m}\Nabla\otimes\vec{P} - \frac{\Eins}{m}(A_1\rho + A_4 \Nabla^2\rho ) 
\notag\\&\ERAI{} -\frac{2A_4}{m}\Nabla\otimes\Nabla\rho \label{dotvp}
\end{align}
with the unit matrix $\Eins$.

Starting from the very general model given by \cref{generalconti,dotP,dotv,dotvp}, various approximations can be made. In most active matter models, it is assumed that the polarization $\vec{P}$ is slow compared to the velocity $\vec{v}$. While this is reasonable for strongly damped systems, $\vec{v}$ should be slow in a system with weak damping and activity because there the momentum density is (almost) a conserved quantity (unlike $\vec{P}$). In this case, it is plausible to assume that $\vec{v}$ evolves slower than $\vec{P}$. This limit, which is less well understood, will be considered in this work. As is shown in detail in \cref{qsa}, the quasi-stationary approximations \cite{CatesT2013,WittkowskiSC2017,BickmannW2020,BickmannW2019b} $\dot{\Tensor{v}}_P=\Tensor{0}$  and $\dot{\vec{P}}=\vec{0}$ allow one to derive
\begin{align}
\dot{\vec{v}} + (\vec{v}\cdot\Nabla) \vec{v} &= -\frac{1}{m}\Nabla(f'(\rho)- (\tilde{\kappa}+\delta \rho) \Nabla^2 \rho + \tilde{\lambda}(\Nabla \rho)^2+U_1) \nonumber
\\&\ERAI{}- \gamma \vec{v} + \frac{v_{\mathrm{ld}}(\rho)^2}{\gamma}\Nabla^2\vec{v} +\xi(\Nabla^2\rho)\Nabla\rho,
\label{dotvgeneral2}
\end{align}
where $\tilde{\kappa}$, $\delta$, and $\tilde{\lambda}$ are constant parameters defined in \cref{tildekappa,delta,tildelambda}. Finally, we separate the variational and non-variational dynamics using an argument adapted from Ref.\ \cite{WittkowskiTSAMC2014}. While $-\rho\Nabla^2\rho$ is non-variational, one could write $-(\rho\Nabla^2\rho+(\Nabla\rho)^2/2)$ as a derivative of the free energy density $\rho(\Nabla\rho)^2/2$. On this basis, we replace $-\delta\rho\Nabla^2\rho$ by $-\delta\rho\Nabla^2\rho-\delta(\Nabla\rho)^2/2+\delta(\Nabla\rho)^2/2$ and combine the last term (i.e., $\delta(\Nabla\rho)^2/2$) with the term $\tilde{\lambda}(\Nabla\rho)^2$ already present to get a term $(\tilde{\lambda}+\delta/2)(\Nabla\rho)^2$. The remaining gradient contribution $- (\tilde{\kappa}+\delta \rho) \Nabla^2 \rho-\delta(\Nabla\rho)^2/2$ can be written as a functional derivative of the passive free energy $F_P = \TINT{}{}{^2r}\kappa(\rho)(\Nabla\rho)^2/2$ with $\kappa(\rho)=\tilde{\kappa}+\delta\rho$. As is standard in passive model B \cite{WittkowskiTSAMC2014}, we make the simplifying assumption that $\kappa$ is constant. Also, we define $\tilde{\lambda}+\delta/2 = \lambda$. Then, \cref{dotvgeneral2} reduces to \cref{dotvgeneral} of AMI+. In the passive limit, we have $v_0=0$, $A_1=0$, and $A_3=A_4 = 0$ since the pair-distribution function is isotropic in the passive case \cite{BialkeLS2013}. In this case, most terms on the right-hand side of \cref{dotvgeneral} vanish, such that it can be written as $-\frac{1}{m}\Nabla\Fdif{F}{\rho}$ with a free energy $F=\TINT{}{}{^2r}(f(\rho)+\rho U_1)$. (Of course, this does not mean that there can be no gradient terms in the passive case. These can be obtained by a more sophisticated treatment of interaction terms.)

\subsection{\label{ddss}Density-dependent swimming speed}
The existence of a density-dependent swimming speed is crucial for the phenomenon of MIPS in active matter. In the case of ABPs, the particles are slowed down by collisions with other particles in a dense phase, which reduces the effective swimming speed in such a phase and creates a feedback that leads to phase separation \cite{CatesT2015}.

One can calculate the density-dependent swimming speed $v_{\mathrm{D}}$ from the interaction-expansion method by looking for a contribution of the form $\Nabla\cdot(v_{\mathrm{D}}[\rho]\uu\varrho)$ in the dynamic equation for $\varrho$ \cite{BickmannW2019b}. Inserting \cref{expansion2} into \cref{dotvarrho} gives
\begin{equation}
\dot{\varrho} = - \Nabla\cdot(\vec{v}\varrho) - \Nabla\cdot(\uu\cdot \Tensor{v}_{\vec{P}}\varrho) + D_R \partial_\phi^2\varrho. 
\end{equation}
This shows that the role of the density-dependent swimming speed is, in our extended theory, played by the tensorial quantity $ \Tensor{v}_{\vec{P}}$. The part of this tensor that is proportional to $\Eins$ then directly gives us $v_{\mathrm{D}}$. As shown in \cref{qsa}, this tensor is given by
\begin{equation}
\begin{split}
\Tensor{v}_{\vec{P}}&=v_{\mathrm{D}}[\rho,\vec{v}]\Eins -\bigg(\frac{2A_4}{\gamma m}-\frac{v_0 k_B T}{\gamma \varrho_0 m D_R}\bigg)\Nabla\otimes\Nabla\rho\\
&\ERAI{}-\frac{k_B TA_1}{\gamma^2 \varrho_0 m^2 D_R}\Nabla\otimes\Nabla\rho^2-\frac{v_{\mathrm{ld}}(\rho)}{\gamma}\Nabla\otimes\vec{v} 
\\&\ERAI{}- \frac{k_B T}{\gamma^2\varrho_0 m D_R}\Nabla\otimes(\Nabla\cdot(v_{\mathrm{ld}}(\rho)\rho\Nabla\otimes\vec{v}))
\\&\ERAI{}+\frac{k_B T A_1}{\gamma^3 \varrho_0 m^2 D_R}\Nabla\otimes\Nabla(\rho((\vec{v}\cdot\Nabla)\rho))
\end{split}
\label{v0eins}
\end{equation}
with
\begin{equation}
v_{\mathrm{D}}[\rho,\vec{v}]= v_{\mathrm{ld}}(\rho) - \frac{A_4}{\gamma m}\Nabla^2\rho +\frac{A_1}{\gamma^2m}(\vec{v}\cdot\Nabla)\rho.
\label{densitydependentswimmingspeed}
\end{equation}
Equation \eqref{densitydependentswimmingspeed} provides a microscopic expression for the density-dependent swimming speed in the active fluid. Interestingly, the \ZT{density-dependent} swimming speed depends not only on the density $\rho$, but also on the velocity $\vec{v}$. This suggests that $\vec{v}$ also has to be taken into account when describing the emergence of MIPS in underdamped active fluids.

\subsection{\label{defi}Mechanical vs thermodynamic velocity}
It is worth briefly discussing here the definition of the velocity field $\vec{v}$. In the theory of classical passive fluids, this can be done in two ways:
\begin{enumerate}
    \item Mechanically: The density $\rho$ obeys a continuity equation $\dot{\rho} = - \Nabla\cdot\vec{g}$ with the momentum density $\vec{g} = \INT{}{}{^2p}\vec{p}P_1(\vec{r},\vec{p})$. We can then define the velocity field as $\vec{v} =\vec{g}/\rho$ \cite{AneroE2007}, implying that
    \begin{equation}
        \dot{\rho} = - \Nabla\cdot(\rho\vec{v}). 
        \label{rhov}
    \end{equation}
    \item Thermodynamically: We assume that the one-body distribution function $P_1$ takes the local equilibrium form
    \begin{equation}
     P_1(\vec{r},\vec{p})\propto\exp\!\bigg(-\frac{(\vec{p}-m\vec{v})^2}{2mk_B T}\bigg)  
     \label{led}
    \end{equation}
    and then \textit{define} the velocity field to be the field $\vec{v}$ appearing in \cref{led} \cite{Grabert1982}. Multiplying the dynamic equation for $P_1$ by $\vec{p}$, inserting \cref{led}, and integrating over $\vec{p}$ also leads to \cref{rhov}.
\end{enumerate}

One might question here whether the thermodynamic definition is sufficiently general as it appears to depend on the fluid being in a local equilibrium state. However, this second definition is actually very general, as can be seen from its application in the Mori-Zwanzig formalism \cite{Mori1965,Zwanzig1960,teVrugtW2019d,teVrugtHW2021,teVrugt2021}. When deriving the equations of hydrodynamics in the Mori-Zwanzig formalism (as done, e.g., in Ref.\ \cite{Grabert1982}), one introduces a \textit{relevant} distribution $\bar{P}_N$ that has the local equilibrium form \cite{WittkowskiLB2012}
\begin{equation} 
\bar{P}_N\propto \exp\!\bigg(-\frac{H-\sum_{i=1}^{w}a_i^\natural A_i}{k_B T}\bigg)
\label{relevantdensity}
\end{equation}
with the Hamiltonian $H$, the thermodynamic conjugates $a_i^\natural$ of the mean values $a_i$ of the relevant variables $A_i$, and the number of relevant variables $w$. The Hamiltonian typically has the form $H=U(\{\vec{r}_i\}) + \sum_{i=1}^{N}\vec{p}_i^2/(2m)$ with the total interaction potential $U$. In fluid mechanics, we use as a relevant variable the total momentum density $\vec{g}=\sum_{i=1}^{N}\vec{p}_i\delta(\vec{r}-\vec{r}_i)$ with the Dirac delta distribution $\delta$. Inserting $\vec{g}$ into \cref{relevantdensity}, writing $\vec{v}$ for $\vec{g}^\natural$, and integrating over the phase-space coordinates of all except for one particle gives
\begin{equation}
\bar{P}_1\propto\exp\!\bigg(-\frac{1}{k_B T}\bigg(\frac{\vec{p}^2}{2m}-\vec{v}\cdot\vec{p}\bigg)\!\bigg),
\label{relevantdensityexample}
\end{equation}
which is proportional to the result \eqref{led}. Thus, the velocity field $\vec{v}$ is simply the thermodynamic conjugate for the momentum density \cite{Grabert1982}. If the relevant (local equilibrium) distribution \eqref{relevantdensityexample} always equals the actual distribution, we have an ideal fluid, otherwise we have dissipation. 

Since both definitions give the same result \eqref{rhov} in the passive case, their difference is usually not even mentioned. However, they are not equivalent for the active fluid considered here. While the mechanical route leads to \cref{rhov} also in the active case, the thermodynamic route considered here gives the different result \eqref{dotrhogeneral}. From the Mori-Zwanzig point of view, our local equilibrium distribution \eqref{localeq} differs from the standard one \eqref{led} due to the presence of additional relevant degrees of freedom. An essential parameter for active phase separation in overdamped \cite{CatesT2015} and underdamped \cite{MandalLL2019} active fluids is the average of $\uu\cdot\vec{p}$ (which corresponds to the average of the projection of the particle momentum onto the direction of self-propulsion). Motivated by this observation, we use the \ZT{momentum density polarization} $\Tensor{g}_{\vec{P}}=\sum_{i=1}^{N}\uu_i\otimes\vec{p}_i\delta(\vec{r}-\vec{r}_i)$ as a relevant variable in addition to $\vec{g}$. Using the same steps as before, \cref{relevantdensity} then gives
\begin{equation}
\bar{P}_1\propto\exp\!\bigg(-\frac{1}{k_B T}\bigg(\frac{\vec{p}^2}{2m}-\vec{v}\cdot\vec{p}-\Tensor{v}_{\vec{P}}:(\uu\otimes\vec{p})\bigg)\!\bigg),
\label{relevantdensityexample2}
\end{equation}
where : denotes a double tensor contraction. Hence, $ \Tensor{v}_{\vec{P}}$ is the thermodynamic conjugate for $\Tensor{g}_{\vec{P}}$. The form \eqref{relevantdensityexample2} is proportional to our local equilibrium form \eqref{localeq}, as can be seen by inserting the expansion \eqref{expansion2} into \cref{localeq}. When turning to a reduced description in terms of $\rho$ and $\vec{v}$ only (namely AMI+), the deviation of \eqref{relevantdensityexample2} from \eqref{relevantdensityexample} gives rise to a \ZT{viscosity term} in \cref{dotvgeneral} with a viscosity proportional to $v_{\mathrm{ld}}(\rho)$, consistent with the fact that deviations from the distribution \eqref{led} give rise to the standard viscosity in a passive fluid \cite{Archer2009}. We are using the thermodynamic route in this work because it makes the microscopic derivation significantly easier as it is more natural from a statistical mechanics point of view. Whichever route we choose, it is a remarkable observation that they do not agree. This result is reminiscent of the one obtained for pressure in the case of AMB (where it is found that the mechanical and the thermodynamic definition no longer coincide for active fluids \cite{WittkowskiTSAMC2014}).

\subsection{\label{schro}Derivation of the Schr\"odinger equation}
Now, we derive the Schr\"odinger equation from AMI given by \cref{activemodelbu1,activemodelbu2}. For this purpose, we assume $f'(\rho)=0$ and $\gamma=0$. Then, \cref{activemodelbu2} reads
\begin{equation}
\dot{\vec{v}} + (\vec{v}\cdot \Nabla) \vec{v} = -\frac{1}{m}\Nabla(-\kappa\Nabla^2 \rho + \lambda(\Nabla \rho)^2+ \Nabla U_1).
\label{activemodelbu3}
\end{equation}
The limit $\gamma\to 0$ has to be taken in a very careful way here to avoid a blow-up resulting from the fact that the system is then active (implying energy influx) but not damped. This problem can be addressed by letting damping and activity go to zero in such a way that the damping term $-\gamma\vec{v}$ vanishes, while the active terms remain finite. In the passive limit, the parameters $A_1$, $A_3$, and $A_4$ will vanish since $g$ has no angular dependence in this case. From the microscopic definitions given by \cref{fmicro,tildekappa,delta,tildelambda,xi}, it can be seen that we can avoid a blow-up by letting $\gamma$, $v_0$, $A_1$, $A_3$, and $A_4$ go to zero at the same rate. In this case, we have $f'=k_B T \ln(\rho) +A_2\rho + (3A_1^2\rho^2)/(4\gamma^2m)$, $\tilde{\kappa}=0$, $\delta = -3A_1A_4/(2\gamma^2m) + 3v_0 A_1 k_B T/(2\gamma^2\varrho_0 m D_R)$, $\tilde{\lambda}=A_1A_4/(2\gamma^2m) - 5 v_0 A_1 k_B T/(4\gamma^2\varrho_0 m D_R)$, and $\xi = A_1 A_4/(\gamma^2m^2) - v_0 A_1 k_B T/(2\gamma^2\varrho_0 m^2 D_R)$. More generally, we can -- as will be discussed later -- still derive an interesting quantum-mechanical model (resembling models used in Refs.\ \cite{Tsekov2009,MoczS2015}) for $\gamma\neq 0$. Thus, here, the assumption $\gamma=0$ is primarily made to simplify the derivation and will be dropped later. Moreover, the static mapping used in \cref{sec:tunnel} can be made already for the overdamped AMB, such that our later analysis of the tunnel effect does not hang on the assumption $\gamma \to 0$.

We define $\rho_q = 2\rho$ ($\rho_q$ will later be interpreted as the quantum-mechanical density) and assume that $\rho_q$ has only small deviations from a spatially and temporally constant reference density $\rho_0$. Noting that adding a constant to $\rho_q$ has no influence on the dynamics once we have set $f'=0$, we can then approximately write
\begin{equation}
\rho=\frac{1}{2}\rho_q = \frac{\rho_0}{2}\frac{\rho_q}{\rho_0}\approx \frac{\rho_0}{2}\ln\!\Big(\frac{\rho_q}{\rho_0}\Big)+\,\text{irrelevant constant}.
\label{substitution}
\end{equation}
As Eq.\ \eqref{activemodelbu1} is linear in $\rho$, it is left unchanged by the replacement $\rho \to \rho_q/2$, i.e., it holds for $\rho_q$ in exactly the same way as for $\rho$. In \cref{activemodelbu3}, we replace $\rho$ by $\rho_0\ln(\rho_q/\rho_0)/2$ (motivated by \cref{substitution}) and assume $\kappa = - \lambda\rho_0$. This gives
\begin{equation}
\begin{split}
\dot{\vec{v}} + (\vec{v} \cdot \Nabla) \vec{v} &= \frac{\kappa\rho_0}{2m}\Nabla\bigg(\Nabla^2 \ln\!\bigg(\frac{\rho_q}{\rho_0}\bigg)+ \frac{1}{2}\bigg(\Nabla \ln\!\bigg(\frac{\rho_q}{\rho_0}\bigg)\! \bigg)^2\bigg)
\\&\ERAI{} -\frac{1}{m}\Nabla U_1
\\ &= \frac{1}{m}\Nabla\bigg(\kappa\rho_0 \frac{\Nabla^2 \sqrt{\rho_q}}{\sqrt{\rho_q}}- U_1\bigg).
\end{split}
\label{activemodelbu4}
\end{equation}
The last step uses the identity \cite{HeifetzC2015}
\begin{equation}
\frac{\Nabla^2 \xi}{\xi} = \Nabla^2\ln(\xi)+(\Nabla\ln(\xi))^2,
\end{equation}
where $\xi$ is a function, and the fact that $\ln(\xi) = 2 \ln(\sqrt{\xi})$. Moreover, we set
\begin{equation}
\frac{\hbar^2}{2m}=\kappa\rho_0   
\label{kappahbar}
\end{equation}
with the reduced Planck constant $\hbar$. We then arrive at the Madelung equations \cite{Madelung1927}
\begin{align}
\dot{\rho}_q &= - \Nabla \cdot (\rho_q \vec{v}),\label{madelung1}\\
\dot{\vec{v}} + (\vec{v}\cdot \Nabla) \vec{v} &=\frac{1}{m}\Nabla\bigg(\frac{\hbar^2}{2m} \frac{\Nabla^2 \sqrt{\rho_q}}{\sqrt{\rho_q}}- U_1\bigg)\label{madelung2}.
\end{align}
Next, we assume that $\vec{v}$ is a potential flow such that we can write
\begin{equation}
\vec{v} = \frac{1}{m}\Nabla S,   
\label{madelung4}
\end{equation}
with a phase $S$, and that $\vec{v}$ satisfies the condition (see Ref.\ \cite{Wallstrom1994})
\begin{equation}
m\CINT{L}{}{\vec{l}}\cdot\vec{v}=2\pi n\hbar
\end{equation}
with a closed loop $L$ and $n\in\mathbb{Z}$. We can then substitute
\begin{equation}
\psi = \sqrt{\rho_q}e^{\frac{\ii}{\hbar}S},\label{madelung3}
\end{equation}
where $\psi$ is (an analogue of) the wave function and $\ii$ is the imaginary unit. Combining \cref{madelung1,madelung2,madelung3,madelung4} then finally yields the Schr\"odinger equation
\begin{equation}
\ii \hbar \partial_t \psi = - \frac{\hbar^2}{2m}\Nabla^2\psi+ U_1\psi.
\label{schroedinger}
\end{equation}

It is also interesting to see what happens if we do not set $f'(\rho_q)=0$. In this case, \cref{activemodelbu4} reads
\begin{equation}
\dot{\vec{v}} + (\vec{v} \cdot \Nabla) \vec{v} = \frac{1}{m}\Nabla \bigg(\kappa\rho_0\frac{\Nabla^2 \sqrt{\rho_q}}{\sqrt{\rho_q}}-f'\bigg(\frac{\rho_q}{2}\bigg)-U_1\bigg).  
\label{nonlinearsgener}
\end{equation}
Applying the substitutions \eqref{kappahbar}, \eqref{madelung4}, and \eqref{madelung3} then gives the nonlinear Schr\"odinger equation \cite{KhesinMM2019}
\begin{equation}
\ii \hbar \partial_t \psi = - \frac{\hbar^2}{2m}\Nabla^2\psi+ U_1\psi + f'\bigg(\frac{|\psi|^2}{2}\bigg)\psi.   
\end{equation}
For example, if we set  $f' = a\rho_q$, we find \cite{MoczS2015}
\begin{equation}
\ii \hbar \partial_t \psi = - \frac{\hbar^2}{2m}\Nabla^2\psi+ U_1\psi + a|\psi|^2\psi,    
\end{equation}
which is the Gross-Pitaevskii equation. This equation has a wide range of applications, such as modeling Bose-Einstein condensates (see Ref.\ \cite{MoczS2015} and references therein).

In the considered active matter system, an even more realistic case would be to have\footnote{From now on, we ignore (as is common) the factors $\Lambda^2$ or $\rho_0$ required to make the argument of the logarithm dimensionless.} $f'=k_B T \ln(\rho_q/2)$ (as suggested by the microscopic derivation in \cref{mikro}) and $\gamma\neq 0$. In this case, \cref{activemodelbu4} reads\footnote{The expressions $\ln(\rho_q/2)$ and $\ln(\rho_q)$ differ only by an additive constant $\ln(2)$, which vanishes after applying $\Nabla$. Therefore, we can simply write $\ln(\rho_q)$ here.}
\begin{equation}
\dot{\vec{v}} + (\vec{v} \cdot \Nabla) \vec{v} = \frac{1}{m}\Nabla\bigg(\kappa\rho_0\frac{\Nabla^2 \sqrt{\rho_q}}{\sqrt{\rho_q}}- k_B T \ln(\rho_q) - U_1 \bigg) - \gamma \vec{v}. 
\label{dissipatives}
\end{equation}
The Madelung transformation then leads to \cite{Tsekov2009}
\begin{equation}
\ii \hbar \partial_t \psi = - \frac{\hbar^2}{2m}\Nabla^2\psi+k_B T \ln(|\psi|^2)\psi + U_1\psi -\frac{\ii\hbar\gamma}{2}\ln\!\bigg(\frac{\psi}{\psi^\star}\bigg)\psi,
\label{dissipativedft}
\end{equation}
which is the Schr\"odinger-Langevin equation \cite{Kostin1972}.
 
From a mathematical perspective, the fact that AMI allows to derive the Madelung equations is essentially a consequence of the fact that AMI is a compressible Euler equation with the pressure being given by the most general expression of a certain order in gradients and densities. Since the Madelung equations are of this order in gradients and densities (if we can approximate densities by their logarithm), they must be contained in AMI. The reason why we use AMI and not the more general AMI+ is that AMI+ contains (some form of) viscosity and leads to velocity fields with non-vanishing rotation, which does not make sense if we want to think of the velocity as the gradient of a phase. It has been shown in the overdamped case that coarse-graining the dynamics of quorum-sensing particles (which have a density-depending swimming speed) leads to active model B rather than B+, i.e., the rotational term is not relevant for these particles \cite{Cates2019}. This indicates that if one wants to study the \ZT{active quantum} effects discussed in the present paper experimentally or via particle-based simulations, quorum-sensing bacteria could be a more promising system than ABPs. What is particularly interesting is that we have found a classical system where the \ZT{quantum potential} from the Madelung equations gives, in a certain limit, an appropriate expression for the classical (nonequilibrium) chemical potential, and that this classical system is active. On the other hand, the derivation shows that AMI provides an approximate description of a quantum system if $f'=0$  and $\gamma=0$ and if $\kappa=-\lambda\rho_0$ is chosen according to \cref{kappahbar} ($\rho_0$ should then be the average probability density of the quantum-mechanical particle).

\subsection{Analogy to dark matter}
An important field of application for the Madelung equations is the study of dark matter \cite{Ferreira2021}. Recently, there has been an increase of interest in so-called \textit{fuzzy dark matter} (FDM), which consists of ultralight scalar particles. Research on FDM is motivated by the lack of evidence for other dark matter candidates, by the fact that such ultralight particles are predicted by various models from particle physics (such as string theory), and by the interesting pattern formation effects that it leads to. The FDM particles are mostly in the ground state and can thus be described by a single macroscopic wavefunction as in a Bose-Einstein condensate \cite{MayS2021}. This wavefunction can (in the nonrelativistic limit without cosmic expansion) be described by the Madelung equations \eqref{madelung1} and \eqref{madelung2} coupled to the Poisson equation \cite{VeltmaatN2016}
\begin{equation}
\Nabla^2 U_1 = 4\pi G m^2\rho_q \label{poisson}
\end{equation}
with the gravitational constant $G$. Equation \eqref{poisson} determines $U_1$, which is here the gravitational potential, via the density $\rho_q$.

Let us now consider the dynamics of a system of ABPs with an electric charge $q$. By the reasoning from \cref{mikro}, its dynamics would be given by AMI in the limit where $v_{\mathrm{ld}}$ and $\xi$ are small. The (electrostatic) potential $U_1$ could be calculated from the charge distribution $\rho$ via the Poisson equation
\begin{equation}
\Nabla^2 U_1 = -\frac{q^2}{\epsilon}\rho\label{poissone}    
\end{equation}
with the permittivity $\epsilon$.

As shown in \cref{schro}, AMI contains the Madelung equations as a limiting case. Therefore, in the \ZT{quantum limit} considered in \cref{schro}, an underdamped charged active matter system would be described by equations of the same form as a fuzzy dark matter system, suggesting an interesting parallel between active and astrophysical systems. Interestingly, also generalized Madelung equations with $f'\neq 0$ and $\gamma \neq 0$ have been used in the context of dark matter physics \cite{MoczS2015}, implying that the analogy between dark and active matter persists also in this more general and more realistic case. The analogy between dark and active matter is further supported by the fact that (as mentioned above) fuzzy dark matter models are based on Bose-Einstein condensates, which have been found also in active matter \cite{MengMMG2021,Golestanian2019,MahaultG2020}.

Note, however, that there is also an important difference, namely the fact that the density appears with a different sign in the gravitational Poisson equation \eqref{poisson} and the electrostatic Poisson equation \eqref{poissone}. This is a consequence of the fact that gravity is a purely attractive force, whereas electrostatic forces are repulsive for particles of the same charge. Therefore, the patterns observed in fuzzy dark matter and in charged active systems might be quite different.

\subsection{\label{sec:tunnel}Tunnel effect}
In this section, we restrict ourselves to one-dimensional systems.

\subsubsection{Tunnel effect in quantum mechanics}
Time-independent problems in quantum mechanics can be described by the stationary Schr\"odinger equation
\begin{equation}
E\psi = -\frac{\hbar^2}{2m}\ppartial{x}\psi + U_1\psi
\label{stationary}
\end{equation}
with the energy $E$. One of the most remarkable phenomena of quantum mechanics is the \textit{tunnel effect}, where a particle travels through a potential barrier that it could not pass through classically. It can be described theoretically by solving \cref{stationary} for the potential
\begin{equation}
U_1(x)=
\begin{cases}
0 & \text{ for }x< -L,\\
V_0 & \text{ for }-L\leq x \leq L,\\
0 & \text{ for }x> L,
\end{cases}
\label{well}
\end{equation}
where $V_0$ is the height and $2L$ the width of the potential barrier.
As is well known, the solution of \cref{stationary} with the potential \eqref{well} is given by 
\begin{equation}
\psi(x)=
\begin{cases}
e^{\ii k x} + R_1 e^{-\ii k x} & \text{ for }x< -L,\\
T_2 e^{-\varkappa x} + R_2 e^{\varkappa x} & \text{ for }-L\leq x \leq L,\\
T_3 e^{\ii k x} & \text{ for }x> L
\end{cases}
\label{tunnelwavefunction}
\end{equation}
with the wavenumbers
\begin{align}
k&=\sqrt{\frac{2mE}{\hbar^2}},\label{k}\\
\varkappa &= \sqrt{\frac{2m(V_0 - E)}{\hbar^2}}, \label{kappa}
\end{align}
the transmission coefficients $T_2$ and $T_3$, and the reflection coefficients $R_1$ and $R_2$. (Explicit expressions for these coefficients are given in Ref.\ \cite{HeifetzP2020}.) The physical interpretation of the solution \eqref{tunnelwavefunction} is that it describes the wavefunction of a particle with energy $E$ that approaches a rectangular potential barrier of height $V_0 >E$ from the left, decays within the barrier, and continues to travel as a wave on the right of the barrier. Since the particle has thereby passed through a barrier that it could not have passed through classically, this phenomenon is known as the \ZT{tunnel effect}. Due to the linearity of \cref{stationary}, another solution is given by
\begin{equation}
\psi(x)=
\begin{cases}
\frac{1}{\sqrt{2}}(e^{\ii k x} + (R_1+ T_3)e^{-\ii k x}) & \text{ for }x< -L,\\
\frac{1}{\sqrt{2}}(T_2+R_2) (e^{-\varkappa x} + e^{\varkappa x}) & \text{ for }-L\leq x \leq L,\\
\frac{1}{\sqrt{2}}(e^{-\ii k x} + (R_1+ T_3)e^{\ii k x}) & \text{ for }x> L,
\end{cases}
\label{tunnelwavefunction2}\raisetag{3em}
\end{equation}
which is simply the superposition of the solution given by \cref{tunnelwavefunction} and the same solution mirror reflected at $x=0$ (corresponding to a particle coming from the right). Such a symmetric tunneling solution has advantages in a numerical treatment (as it allows to use periodic boundary conditions) and captures the same physics. The quantum-mechanical density $\rho_q = |\psi|^2$ for the solution \eqref{tunnelwavefunction2} is given by
\begin{equation}
\rho_q(x)=
\begin{cases}
\frac{1}{2}(1 + R^2 + 2 R\cos(2kx-\alpha))& \text{ for }x< -L,\\
|T_2 + R_2|^2(\cosh(2\varkappa x)+1) & \text{ for }-L\leq x \leq L,\\
\frac{1}{2}(1 + R^2 + 2 R\cos(2kx + \alpha)) & \text{ for }x> L,
\end{cases}
\label{tunneldensity2}
\end{equation}
where we have written $R_1 + T_3 = R e^{\ii\alpha}$ with the modulus $R$ and the phase $\alpha$ of the complex number $R_1 + T_3$.

Using the Madelung transformation, \cref{stationary} can be rewritten as \cite{HeifetzP2020}
\begin{align}
0&=\partial_x(\rho_q v),\\
E &=\frac{m}{2}v^2-\frac{\hbar^2}{2m}\bigg(\frac{1}{2}\ppartial{x} \ln(\rho_q) +\frac{1}{4}(\partial_x\ln(\rho_q))^2\bigg) + U_1\label{madelungstationary}.
\end{align}
The tunnel effect was studied in a Madelung framework in Ref.\ \cite{HeifetzP2020}, where it was shown that, from a hydrodynamic point of view, it can be seen as arising from a pressure jump at the boundary of the barrier that balances the jump in the potential.

\subsubsection{Tunnel effect in active matter}
We now show how an analogon of the tunnel effect can be found in active matter. For simplicity, we set $v=0$, such that AMI reduces to $\mu =$ const.\ with $\mu$ given by \cref{static}. Solutions of \cref{static} with $\mu =$ const.\ are also stationary solutions of AMB, such that all of the following considerations apply to both AMI and AMB. Defining $\rho = \rho_0\tilde{\rho}$, $\mu =\mu_0\tilde{\mu}$, $f' = \mu_0\tilde{f}'$, $\tilde{U}_1= \mu_0 U_1$, and $x = x_0 \tilde{x}$ (the tildes denote dimensionless quantities and $\rho_0$, $\mu_0$, and $x_0$ are constants) gives
\begin{equation}
\tilde{\mu} = \tilde{f}'(\tilde{\rho}) - \tilde{\kappa} \ppartial{\tilde{x}} \tilde{\rho} +\tilde{\lambda} (\partial_{\tilde{x}}\tilde{\rho})^2 + \tilde{U}_1  
\end{equation}
with $\tilde{\kappa} = \frac{\kappa\rho_0}{x_0^2\mu_0}$ and $\tilde{\lambda} =\frac{\lambda\rho_0^2}{x_0^2\mu_0}$. Dropping all tildes results in
\begin{equation}
\mu = f'(\rho)-\kappa\ppartial{x}\rho + \lambda(\partial_x\rho)^2 + U_1,   
\label{aqmgen} 
\end{equation}
which is simply \cref{static} in dimensionless form. We could have eliminated the parameters $\kappa$ and $\lambda$ by an appropriate choice of the constants $\rho_0$, $\mu_0$, and $x_0$, but we assume here that these constants have already been used to eliminate other parameters, e.g., in the free energy.

We now consider the special case with $\kappa = -\lambda$ and $f'=0$, in which \cref{aqmgen} reduces to
\begin{equation}
\mu = -\kappa(\ppartial{x}\rho + (\partial_x\rho)^2) + U_1,   
\label{aqm} 
\end{equation}
A solution of \cref{aqm} for the potential \eqref{well} is given by
\begin{equation}
\rho(x)=
\begin{cases}
\ln(\cos(k(x+L)+\alpha)) + A & \text{ for }x< -L,\\
\ln(\cosh(\varkappa x)) + B & \text{ for }-L\leq x \leq L,\\
\ln(\cos(k(x-L)-\alpha))+A & \text{ for }x> L
\end{cases}
\label{wellsolution}
\end{equation}
with the wavenumbers
\begin{align}
k &=\sqrt{\frac{\mu}{\kappa}},\label{k2}\\
\varkappa &= \sqrt{\frac{V_0 - \mu}{\kappa}}\label{kappa2},
\end{align}
the phase shift
\begin{equation}
\alpha = \arctan\!\bigg(\frac{\varkappa}{k}\tanh(\varkappa L)\bigg),  
\label{constraint1}
\end{equation}
and two constants $A$ and $B$ that satisfy
\begin{equation}
A-B=\ln(\cosh(\varkappa L))- \ln(\cos(\alpha)).
\label{constraint2}
\end{equation}
Equations \eqref{constraint1} and \eqref{constraint2} ensure that $\rho$ and $\partial_x\rho$ are continuous at the boundaries of the potential barrier. We have thus found an analytical solution of \cref{aqm} for an active system at a potential barrier, namely \cref{wellsolution}. At the boundary of the potential barrier, a discontinuity in $\kappa \ppartial{x}\rho$ balances the discontinuity in $U$ and thereby ensures that $\mu$ is constant (as required for a stationary solution). Following the analysis in Ref.\ \cite{HeifetzP2020}, we can define $I=\lambda(\partial_x\rho)^2$ (which is here the active term) and a pressure $\Pi = -\rho \kappa\ppartial{x}\rho$. $\Pi$ is the pressure one would get from the thermodynamic expression $\mu\rho - f$ for $\lambda=0$ and $f'=0$ \cite{WittkowskiTSAMC2014}. (Here, we have $\lambda\neq 0$, so $\Pi$ is in general not equal to the thermodynamic or mechanical pressure in the active system.) With these definitions, \cref{aqmgen} gives for $f'=0$
\begin{equation}
\mu = \frac{\Pi}{\rho} + I + U_1.
\end{equation}
At the boundaries of the potential barrier, $\rho$ and $I$ are continuous. The \ZT{tunneling} is thus a consequence of a pressure jump $\Delta \Pi = \rho V_0$ at the boundaries.

For $v=0$, the stationary Madelung equation \eqref{madelungstationary} coincides with the stationary form of AMI or AMB given by \cref{aqm} if we identify $\rho=\frac{1}{2}\ln(\rho_q)$ (cf.\ \cref{substitution}), $\kappa=\hbar^2/(2m)$ (cf.\ \cref{kappahbar}), and $\mu = E$. Therefore, we do not even need to employ the approximation \eqref{substitution} from the dynamical case, we can just straightforwardly map the quantum onto the classical problem.\footnote{We are still making an approximation since $v$ is not zero for the tunnel effect \cite{HeifetzP2020}. Nevertheless, the essential physics can still be captured in the simpler case $v=0$.} Taking the logarithm of the quantum solution \eqref{tunneldensity2} does indeed give us something that (apart from phases and prefactors) looks like \cref{wellsolution}, indicating that similar physical mechanisms act here. In particular, the change in the potential leads to a shift in the wavenumber from $k$ to $\varkappa$ that gives a density decay within the barrier for $\mu < V_0$ (or $E<0$) both in the quantum and in the active case (compare \cref{k,kappa} to \cref{k2,kappa2}).

\subsubsection{Numerical continuation}
The strong mathematical analogy between AMI and the Madelung equations (or between AMB and the stationary Schr\"odinger equation) holds only for the rather special case $f'(\rho)=0$ and $\kappa = -\lambda$. In a real experiment, these equalities will, of course, be realized at most approximately. Therefore, it is interesting to investigate how robust the analogy between active matter and quantum mechanics is if these equalities are modified. For this purpose, we consider the more general model
\begin{equation}
\mu = a \rho - \kappa \ppartial{x} \rho + \lambda (\partial_x\rho)^2 + U_1,
\label{moregeneralmodle}
\end{equation}
where $U_1$ is still given by \cref{well}. For $a=0$ and $\kappa=-\lambda$, the analytical solution \eqref{wellsolution} is known. Starting from these parameter values and this solution, we can find solutions for \cref{moregeneralmodle} for other parameter values via numerical continuation (see \cref{cont}). 
 
We wish to ensure that the density $\rho$ is always positive and that $\rho$ and $\partial_x\rho$ take identical values on both boundaries of the domain, allowing us to use periodic boundary conditions. This determines the one-dimensional domain $\Omega=[-\alpha/k-L,L+\alpha/k]$. Furthermore, we set $\kappa =1$, $V_0 =5$, $\mu =1$, and $\alpha = \pi/4$ and use the analytical solution \eqref{wellsolution} as starting solution for the continuation. Using \cref{k2,kappa2,constraint1} we obtain $L=\arctanh(1/2)/2$. Moreover, we set $B=0.5$ (an arbitrary positive constant can be chosen here); $A$ is then determined by \cref{constraint2}. Note that with this also the averaged rescaled particle density is determined as $\bar{\rho}=\TINT{\Omega}{}{x}\rho(x)/|\Omega|\approx 0.7945$ (where $|\Omega|$ is the domain length). It can be chosen arbitrarily using different values of $B$. Hence, the following result does not depend on the particle number. The starting state is now continued, changing various parameters while keeping $\bar{\rho}$ fixed. This in turn determines $\mu$ as corresponding Lagrange multiplier. Alternatively, one could keep $\mu$ fix, in which case $\bar{\rho}$ would change during the continuation. However, this is not pursued here.

\begin{figure*}
\centering
\includegraphics[width=\linewidth]{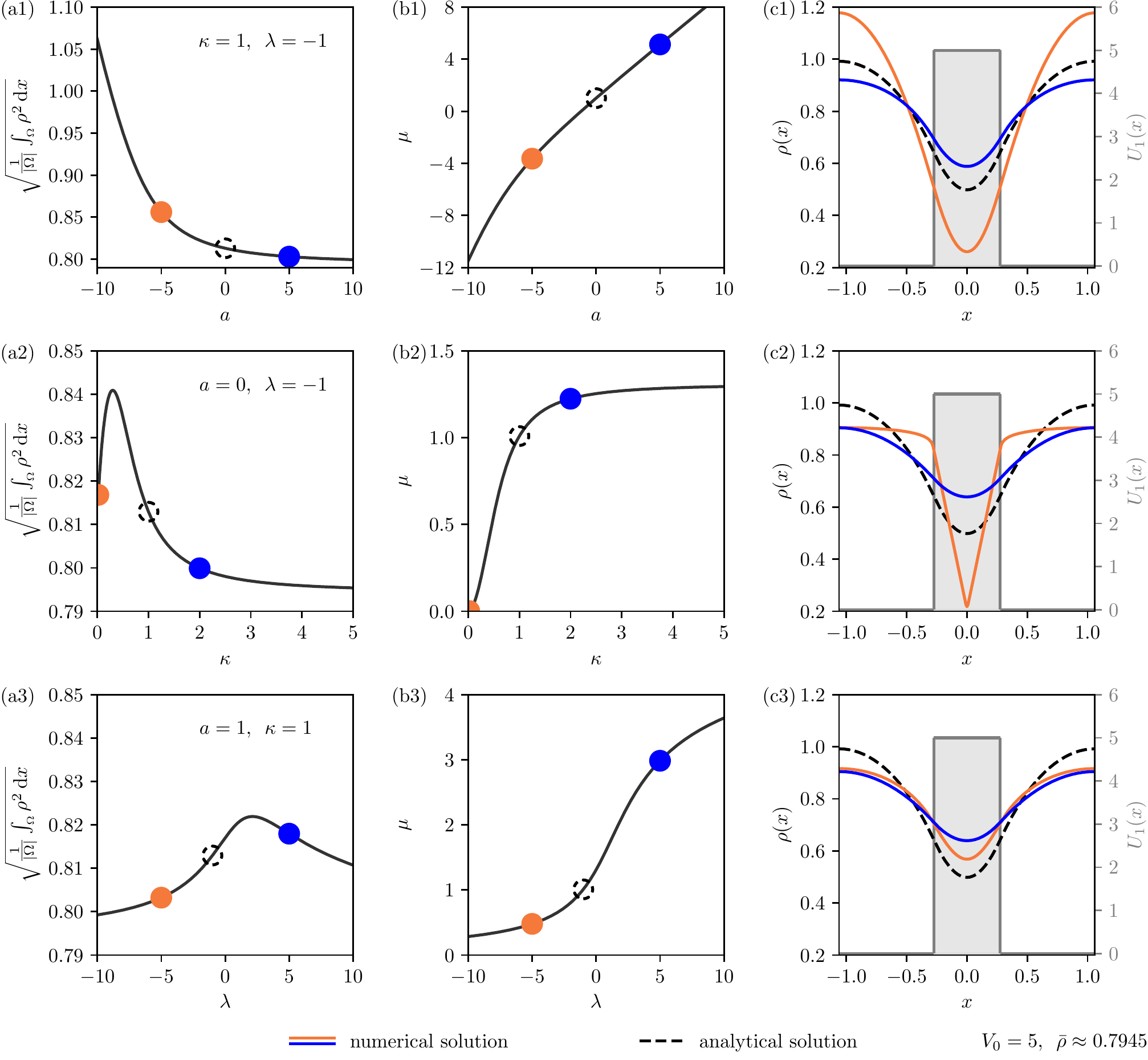}
\caption{\label{fig:tunnel}Results of the numerical continuation of \cref{moregeneralmodle}. Left column: L$_2$-norm of $\rho$ as a function of (a1) $a$, (a2) $\kappa$, and (a3) $\lambda$. Middle column: Chemical potential $\mu$ as a function of (b1) $a$, (b2) $\kappa$, and (b3) $\lambda$. Right column (c1)-(c3): Density profiles for selected parameter values as indicated by circles of corresponding colors in columns (a) and (b) on the domain $\Omega \approx \left[-1.06,1.06\right]$. They are compared to the analytical solution \eqref{wellsolution} (dashed curves), which is used as the starting point for the continuation. If not varied, parameters are given by $\kappa = -\lambda = 1$ and $a = 0$, and we always have $V_0 = 5$ and $\bar{\rho}\approx 0.7945$. Note that the general form of the analytical solution persists also for other parameter values, indicating that the tunnel effect in model \eqref{moregeneralmodle} is a robust phenomenon.}
\end{figure*}

Figure \ref{fig:tunnel} shows bifurcation diagrams and solution profiles that illustrate the tunnel effect that can be observed in model \eqref{moregeneralmodle}. In the left and center columns of Fig.~\ref{fig:tunnel}, we see how the L$_2$-norm $\sqrt{\TINT{\Omega}{}{x}\rho^2(x)/|\Omega|}$ and the generalized chemical potential $\mu$, respectively, depend on the parameters (top) $a$, (center) $\kappa$, and (bottom) $\lambda$.  Finally, the right column of Fig.~\ref{fig:tunnel} presents solution profiles for the states indicated by orange and blue circles in the corresponding bifurcation diagrams to their left. The dashed black curve in each solution plot indicates the analytical solution given by Eq.~\eqref{wellsolution} for comparison. 
 
In general, the plots in the right column of Fig.~\ref{fig:tunnel} show that the general form of the solution does not change significantly if the parameters' values are not exactly those used for the analytical mapping. This indicates that the tunnel effect, and the general active-quantum analogy presented here, are not an artifact of picking the parameter values in such a way that it works, but rather a robust phenomenon that can be investigated also in microscopic simulations and experiments. Furthermore, according to a linear stability analysis that is performed during the continuation (see Section \ref{cont}), the solution is linearly stable with respect to perturbations compatible with mass conservation for all considered parameter values. Despite this limitation and the fact that we consider a small domain, the stability of all solutions emphasizes the relevance for experiments. 

We can also get a more detailed idea of the effect that changing the various parameters has on the solution \eqref{wellsolution}. In general, a steep decrease of $\rho$ towards $x=0$ indicates that the field cannot penetrate far into the potential barrier, whereas a more flat curve is a sign of a strong tunnel effect. Changing $\lambda$ has no strong effect on the form of the solution (Fig.~\ref{fig:tunnel}(c3)). The tunnel effect becomes more pronounced for positive values of $a$, whereas it is suppressed by negative ones (Fig.~\ref{fig:tunnel}(c1)). Since positive values of $a$ are more plausible on physical grounds (one would typically expand $f$ around a local minimum rather than around a local maximum), we can expect this \ZT{tunneling} to be even more significant in real systems. Note that for sufficiently large values of $a$, we get $\mu > V_0$, such that strictly speaking we do not have tunneling anymore (since tunneling requires $E<V_0$ and $\mu$ corresponds to $E$). For $\mu > V_0$, $\varkappa$ becomes imaginary (see \cref{kappa2}) such that $\rho$ has the form $\ln(\cos(x))$ also within the barrier. The strongest effect can be found by varying $\kappa$ (Fig.~\ref{fig:tunnel}(c2)). If it is small (close to zero), we observe a sharp decrease and thus very weak tunneling. For larger $\kappa$, on the other hand, the field can pass through the barrier much more easily. This result is plausible since, as indicated above, it is the discontinuity in the $\kappa$ term that balances the discontinuity of the potential. Also, larger values of $\kappa$ imply that gradients, which are smaller if the fluid passes through the barrier (i.e., if the tunnel effect is present), are associated with an energetic cost, implying that \ZT{tunneling} is more likely to occur for larger $\kappa$. 

\section{\label{conclusion}Conclusions}
In this work, we have systematically derived an extension of common scalar active matter models to the underdamped case which we refer to as \textit{active model I+}. This model and its derivation reveal some interesting and novel features of inertial active matter, such as the fact that mechanical and thermodynamic definitions of the velocity give different results and that the particles' density-dependent swimming speed acts as an effective viscosity. Moreover, we have shown that AMI+ contains (a nonlinear extension of) the Madelung equations and therefore the (nonlinear) Schr\"odinger equation as a special case, such that the Schr\"odinger equation can be seen as an active field theory. This allows to study quantum effects in active-matter systems, as has been demonstrated for the tunnel effect and for fuzzy dark matter. A numerical investigation of the active tunnel effect shows that this active-quantum analogy has no sensitive dependence on the assumptions that have been made to derive it, indicating that it is of broader relevance for both theory and experiment.

\section{Methods}
\subsection{\label{qsa}Quasi-stationary approximation}
Here, we explain in more detail the microscopic derivation of \cref{dotvgeneral2}. Using the quasi-stationary approximation
\begin{equation}
\dot{\Tensor{v}}_{\vec{P}}=\Tensor{0}, 
\label{oversimple}
\end{equation}
\cref{dotvp} gives
\begin{equation}
\begin{split}
\Tensor{v}_{\vec{P}}&=\bigg(v_{\mathrm{ld}}(\rho) - \frac{A_4}{\gamma m}\Nabla^2\rho\bigg)\Eins+ D_R \frac{\vec{v}\otimes\vec{P}}{\gamma\rho}
\\&\ERAI{}-\frac{2A_4}{\gamma m}\Nabla\otimes\Nabla\rho-\frac{k_B T}{\gamma \varrho_0 m}\Nabla\otimes\vec{P}\\
&\ERAI{}-\frac{1}{\gamma}( \Tensor{v}_{\vec{P}}\cdot\Nabla)\otimes\vec{v} -\frac{1}{\gamma}(\vec{v}\cdot\Nabla) \Tensor{v}_{\vec{P}},
\end{split}
\label{v0einsaa}
\end{equation}
where $v_{\mathrm{ld}}$ is defined in \cref{vdrho}. By inserting \cref{v0einsaa} recursively into itself and neglecting terms of higher than second order in gradients, of higher than first order in velocities, or that involve products of polarizations with velocities, we find
\begin{align}
\Tensor{v}_{\vec{P}}&=\bigg(v_{\mathrm{ld}}(\rho) - \frac{A_4}{\gamma m}\Nabla^2\rho\bigg)\Eins-\frac{2A_4}{\gamma m}\Nabla\otimes\Nabla\rho
\label{v0einsa}\\&\ERAI{} -\frac{k_B T}{\gamma \varrho_0 m}\Nabla\otimes\vec{P} 
-\frac{v_{\mathrm{ld}}(\rho)}{\gamma}\Nabla\otimes\vec{v} 
+\frac{A_1}{\gamma^2m}(\vec{v}\cdot\Nabla)\rho\Eins. 
\nonumber
\end{align}
The motivation behind these approximations is that we wish to derive a theory of third order in gradients and of second order in velocities and that we assume both polarizations and velocities to be small. (By \ZT{velocity}, we mean $\vec{v}$, whereas $\Tensor{v}_{\vec{P}}$ is always referred to as \ZT{velocity polarization}.) The velocity polarization $\Tensor{v}_{\vec{P}}$ appears in \cref{dotv} only in the term $ (\Tensor{v}_{\vec{P}}\cdot\Nabla)\cdot\Tensor{v}_{\vec{P}}/2$ (quadratic in $\Tensor{v}_{\vec{P}}$ and of first order in gradients) and in the term $D_R \vec{P}\cdot\Tensor{v}_{\vec{P}}/(2\rho)$ (product with the small polarization). If we insert \cref{v0einsa} into \cref{dotP} and drop again terms containing products of velocities with polarizations (in particular the advection term), we find
\begin{align}
\dot{\vec{P}}&= - \Nabla(\rho v_{\mathrm{ld}}(\rho)) + \frac{A_4}{\gamma m} 
\big(\Nabla(\Nabla\rho)^2+3\rho\Nabla\Nabla^2\rho+(\Nabla\rho)(\Nabla^2\rho)\big)
\nonumber\\&\ERAI{} + \frac{k_B T}{\gamma \varrho_0 m}\Nabla\cdot(\rho\Nabla\otimes\vec{P}) +\Nabla\cdot\bigg( \frac{v_{\mathrm{ld}}(\rho)}{\gamma}\rho\Nabla\otimes\vec{v}\bigg)\nonumber\\&\ERAI{}-\frac{A_1}{\gamma^2m}\Nabla(\rho((\vec{v}\cdot\Nabla)\rho)) - D_R \vec{P},  
\label{dotpa}\raisetag{4ex}
\end{align}
where we used the vector identity
\begin{equation}
\Nabla\cdot(\rho\Nabla\otimes\Nabla\rho)= \frac{1}{2}\Nabla(\Nabla\rho)^2 + \rho\Nabla\Nabla^2\rho. 
\label{vectoridentity}
\end{equation}
If we have $v_{\mathrm{ld}}(\rho)\approx v_0$, the first term on the right-hand side of \cref{dotpa} reduces to the self-propulsion term known from the active PFC model \cite{MenzelL2013}. We now make the further quasi-stationary approximation $\dot{\vec{P}} =\vec{0}$ and find
\begin{align}
\nonumber\vec{P}&= - \frac{1}{D_R}\Nabla(\rho v_{\mathrm{ld}}(\rho))+ \Nabla\cdot\bigg( \frac{v_{\mathrm{ld}}(\rho)}{\gamma D_R}\rho\Nabla\otimes\vec{v}\bigg)\\&\ERAI{} + \frac{A_4}{\gamma m D_R}\big(\Nabla(\Nabla\rho)^2+3\rho\Nabla\Nabla^2\rho+(\Nabla\rho)(\Nabla^2\rho)\big) \label{dotpan}
\\\nonumber&\ERAI{} + \frac{k_B T}{\gamma \varrho_0 m D_R}\Nabla\cdot(\rho\Nabla\otimes\vec{P})-\frac{A_1}{\gamma^2m D_R}\Nabla(\rho((\vec{v}\cdot\Nabla)\rho)).    
\end{align}
Inserting \cref{dotpan} into itself and neglecting terms of higher than third order in gradients and second order in densities gives
\begin{equation}
\begin{split}
\vec{P}&= - \frac{1}{D_R}\Nabla(\rho v_{\mathrm{ld}}(\rho))+ \Nabla\cdot\bigg( \frac{v_{\mathrm{ld}}(\rho)}{\gamma D_R}\rho\Nabla\otimes\vec{v}\bigg)\\&\ERAI{} + \frac{A_4}{\gamma m D_R}\big(\Nabla(\Nabla\rho)^2+3\rho\Nabla\Nabla^2\rho+(\Nabla\rho)(\Nabla^2\rho)\big)
\\&\ERAI{} - \frac{v_0 k_B T}{2\gamma \varrho_0 m D_R^2}(\Nabla(\Nabla\rho)^2 + 2\rho\Nabla\Nabla^2\rho)
\\&\ERAI{}-\frac{A_1}{\gamma^2m D_R}\Nabla(\rho((\vec{v}\cdot\Nabla)\rho)),
\end{split}
\label{qs2}
\end{equation}
where we have used \cref{vectoridentity} again.
This agrees with the result from Ref.\ \cite{BialkeLS2013} if we neglect terms of higher order in gradients and the velocity term in \cref{qs2}. We can insert \cref{qs2} into \cref{v0einsa} and neglect terms of higher than third order in gradients to get \cref{v0eins,densitydependentswimmingspeed}.

Inserting \cref{v0eins,densitydependentswimmingspeed,qs2} into \cref{generalconti} and neglecting terms of higher than second order in gradients gives \cref{dotrhogeneral}. The reason that terms of second order in gradients are sufficient is that all third-order terms would include also $\vec{v}$, which (as is evident from \cref{dotvgeneral}) is of at least first order in gradients.

Deriving \cref{dotvgeneral} is slightly more involved. First, we deal with the term $D_R\vec{P}\cdot\Tensor{v}_{\vec{P}}/(2\rho)$ appearing in \cref{dotv}. Inserting \cref{v0eins,densitydependentswimmingspeed,qs2}, dropping terms of higher than third order in gradients, terms quadratic in $\vec{v}$ that are of higher than first order in gradients (since $\vec{v}$ is of first order in gradients), terms of higher than second order in fields, and products of density gradients and velocities (these approximations will be referred to as \ZT{standard approximations} from here on) gives
\begin{align}
&\frac{D_R\vec{P}\cdot\Tensor{v}_{\vec{P}}}{2\rho}
\nonumber\\&= - \frac{v_0^2\Nabla \rho}{2\rho} + \frac{3v_0 A_1\Nabla\rho}{2\gamma m} 
- \frac{A_1^2}{2\gamma^2m^2}\Nabla\rho^2
\nonumber\\&\ERAI{} + \frac{v_0}{4\rho}\bigg(\frac{2A_4}{\gamma m}-\frac{v_0 k_B T}{\gamma \varrho_0 m D_R}\bigg)\Nabla(\Nabla\rho)^2+ \frac{v_0A_4}{2\gamma m \rho}(\Nabla\rho)(\Nabla^2\rho)
\nonumber\\&\ERAI{}+ \frac{v_0A_4}{2\gamma m \rho}\big(\Nabla(\Nabla\rho)^2+3\rho\Nabla\Nabla^2\rho+(\Nabla\rho)(\Nabla^2\rho)\big) \label{drterm}
\\&\ERAI{} - \frac{v_0^2 k_B T}{4\gamma \varrho_0 m D_R \rho}(\Nabla(\Nabla\rho)^2 + 2\rho\Nabla\Nabla^2\rho)+\frac{v_{\mathrm{ld}}^2(\rho)}{2\gamma}\Nabla^2\vec{v},
\nonumber
\end{align}
where we have used
\begin{equation}
(\Nabla\otimes\Nabla\rho)\cdot\Nabla\rho = \frac{1}{2}\Nabla(\Nabla\rho)^2.\label{iden}
\end{equation}
We have not expanded the expression $v_{\mathrm{ld}}(\rho)^2$ in the last term of \cref{drterm} to simplify the notation even though this term thereby contains terms up to third order in fields. The first term on the right-hand side of \cref{drterm} can be rewritten using $(\Nabla\rho)/\rho = \Nabla\ln(\rho)$. In the fourth-from-last, third-from-last, and penultimate terms, we replace $\rho$ by $\varrho_0$ in the denominator such that these terms are of second order in $\rho$ as required. This yields
\begin{equation}
\begin{split}
\frac{\vec{P}\cdot\Tensor{v}_{\vec{P}}}{2\rho}&= - \frac{v_0^2}{2}\Nabla\ln(\rho) + \frac{3v_0 A_1}{2\gamma m}\Nabla\rho - \frac{A_1^2}{2\gamma^2m^2}\Nabla\rho^2\\
&\ERAI{}+ \bigg(\frac{3v_0A_4}{2\gamma m}- \frac{v_0^2 k_B T}{2\gamma \varrho_0 m D_R}\bigg)\Nabla\Nabla^2\rho\\
&\ERAI{}+\bigg(\frac{v_0A_4}{\gamma m \varrho_0}-\frac{v_0^2k_B T}{2\gamma \varrho_0^2 m D_R}\bigg)\Nabla(\Nabla\rho)^2\\
&\ERAI{}+\frac{v_0A_4}{\gamma m \varrho_0}(\Nabla\rho)(\Nabla^2\rho)+\frac{v_{\mathrm{ld}}^2(\rho)}{2\gamma}\Nabla^2\vec{v}.
\label{drterm2}
\end{split}    
\end{equation}
Next, we consider the term $( \Tensor{v}_{\vec{P}}\cdot\Nabla)\cdot \Tensor{v}_{\vec{P}}/2$. Inserting \cref{vdrho,v0eins,densitydependentswimmingspeed} gives with the standard approximations
\begin{equation}
\begin{split}
\frac{1}{2}( \Tensor{v}_{\vec{P}}\cdot\Nabla)\cdot \Tensor{v}_{\vec{P}}
&= - \frac{v_0 A_1}{2\gamma m}\Nabla\rho + \frac{A_1^2}{4\gamma^2m^2}\Nabla\rho^2 
\\&\ERAI{}-\bigg(\frac{3v_0A_4}{2\gamma m}-\frac{v_0^2 k_B T}{2\gamma \varrho_0 m D_R}\bigg)\Nabla\Nabla^2\rho\\
&\ERAI{}+\bigg(\frac{3A_1 A_4}{2\gamma^2 m^2}-\frac{3v_0A_1 k_B T}{2\gamma^2 \varrho_0 m^2 D_R}\bigg)\Nabla(\rho\Nabla^2\rho)\\
&\ERAI{}+\bigg(\frac{A_1 A_4}{2\gamma^2 m^2}-\frac{5 v_0A_1 k_B T}{4\gamma^2 \varrho_0 m^2 D_R}\bigg)\Nabla(\Nabla\rho)^2
\\
&\ERAI{}- \bigg(\frac{A_1 A_4}{\gamma^2 m^2}-\frac{v_0A_1 k_B T}{2\gamma^2 \varrho_0 m^2 D_R}\bigg)(\Nabla\rho)(\Nabla^2\rho) 
\\&\ERAI{}- \frac{v_{\mathrm{ld}}(\rho)^2}{2\gamma}\Nabla^2\vec{v},
\end{split}    
\label{vterm}\raisetag{2em}
\end{equation}
where we used \cref{iden} and
\begin{align}
\Nabla^2\rho^2 &=2(\rho\Nabla^2\rho + (\Nabla\rho)^2),\label{identitaet}\\
\rho\Nabla\Nabla^2\rho &=\Nabla(\rho\Nabla^2\rho)- (\Nabla\rho)(\Nabla^2\rho).
\end{align}
Finally, using \cref{qs2,identitaet} and the standard approximations, we find
\begin{equation}
\begin{split}
&A_3\Nabla^2\vec{P}+2A_3\Nabla(\Nabla\cdot\vec{P}) 
\\&= -\frac{3v_0A_3}{D_R}\Nabla\Nabla^2\rho + \frac{6A_1A_3}{\gamma m D_R}\Nabla(\rho\Nabla^2\rho+(\Nabla\rho)^2).
\end{split}
\label{pterms}
\end{equation}
Inserting \cref{vterm,pterms,drterm2} into \cref{dotv} and collecting terms results in
\begin{align}
\dot{\vec{v}}&= - (\vec{v}\cdot\Nabla)\vec{v} -\gamma \vec{v}\nonumber
\\&\ERAI{}-\frac{1}{m}\Nabla \bigg(\!\bigg(k_B T+\frac{mv_0^2}{2}\bigg)\ln(\rho) \nonumber
\\&\ERAI{} +\bigg(A_2-\frac{2v_0A_1}{\gamma}\bigg) \rho
+\frac{3A_1^2}{4\gamma^2m} \rho^2 \nonumber\\&\ERAI{}
-\bigg(\frac{3v_0A_3}{D_R}+\frac{3v_0A_4}{\gamma }-\frac{v_0^2 k_B T}{\gamma \varrho_0 D_R}\bigg)\Nabla^2\rho\nonumber
\\&\ERAI{}-\bigg(-\frac{6A_1A_3}{\gamma m D_R}-\frac{3A_1 A_4}{2\gamma^2 m}+\frac{3v_0A_1 k_B T}{2\gamma^2 \varrho_0 m D_R}\bigg)(\rho\Nabla^2\rho)\nonumber
\\&\ERAI{}+\bigg(-\frac{v_0A_4}{\gamma\varrho_0}+\frac{v_0^2k_B T}{2\gamma\varrho_0^2D_R}+\frac{A_1A_4}{2\gamma^2m}\label{dotvcomplex2}
\\&\ERAI{}-\frac{5v_0A_1k_B T}{4\gamma^2\varrho_0mD_R}+\frac{6A_1A_3}{\gamma m D_R}\bigg)(\Nabla\rho)^2 +U_1\bigg)\nonumber
\\&\ERAI{}+\bigg(\frac{A_1 A_4}{\gamma^2 m^2}-\frac{v_0A_1 k_B T}{2\gamma^2 \varrho_0 m^2 D_R} +\frac{v_0A_4}{\gamma m \varrho_0}\bigg)(\Nabla\rho)(\Nabla^2\rho)\nonumber 
\\&\ERAI{}+ \frac{v_{\mathrm{ld}}(\rho)^2}{\gamma}\Nabla^2\vec{v}.\nonumber
\end{align}
We have not dropped the higher-order contributions in $\rho$ for the logarithmic term, which is consistent with the fact that we do not make a constant-mobility approximation \cite{ArcherRRS2019}. It is interesting that, instead of the thermal energy $k_B T$ we would have in the passive case, the ideal gas contribution $f'$ is proportional to $k_B T + mv_0^2/2$, implying that the active contribution to the kinetic energy effectively shifts the temperature by $mv_0^2/(2k_B)$. This is a different sort of \ZT{effective temperature} than the one reported for active systems in Refs.\ \cite{PreislerD2016,BickmannBJW2020}. The additional term $m v_0^2 \ln(\rho)/2$ originates from \cref{drterm2}. Equation \eqref{dotvcomplex2} has the form of \cref{dotvgeneral2}, such that we can read off the expressions
\begin{align}
\begin{split}
f'(\rho)&=\bigg(k_B T+\frac{mv_0^2}{2}\bigg)\ln(\rho) 
\\&\ERAI{} +\bigg(A_2-\frac{2v_0A_1}{\gamma}\bigg) \rho
+\bigg(\frac{3A_1^2}{4\gamma^2m} \bigg)\rho^2,\label{fmicro}
\end{split}\\
\tilde{\kappa} &= \frac{3v_0A_3}{D_R}+\frac{3v_0A_4}{\gamma }-\frac{v_0^2 k_B T }{\gamma \varrho_0 D_R},\label{tildekappa}
\\
\delta &=-\frac{6A_1A_3}{\gamma m D_R}-\frac{3A_1 A_4}{2\gamma^2 m}+\frac{3v_0A_1 k_B T}{2\gamma^2 \varrho_0 m D_R}, \label{delta}
\\\begin{split}
\tilde{\lambda}&=-\frac{v_0A_4}{\gamma\varrho_0}+\frac{v_0^2k_B T}{2\gamma\varrho_0^2D_R}+\frac{A_1A_4}{2\gamma^2m}
\\&\ERAI{}-\frac{5v_0A_1k_B T}{4\gamma^2\varrho_0mD_R}+\frac{6A_1A_3}{\gamma m D_R},\label{tildelambda}
\end{split}\\
\xi &=\frac{A_1 A_4}{\gamma^2 m^2}-\frac{v_0A_1 k_B T}{2\gamma^2 \varrho_0 m^2 D_R} +\frac{v_0A_4}{\gamma m \varrho_0}.
\label{xi}
\end{align}

\subsection{\label{cont}Numerical path continuation}
In Section~\ref{sec:tunnel}, we use numerical path continuation via the \textit{Matlab} package \textit{pde2path}~\cite{UeckerWR2014}. 
Starting from the analytical solution (Eq.~\eqref{wellsolution}) of model \eqref{moregeneralmodle}, \textit{pde2path} subsequently applies tangent predictors and Newton correctors to track a branch of steady states through parameter space. A numerical linear stability analysis during the continuation yields the stability of the corresponding solution and enables the detection of bifurcations. \textit{pde2path} uses the finite element method and the model is implemented in a weak formulation. We have used a primary control parameter ($a$, $\kappa$, or $\lambda$) and the chemical potential $\mu$ as a secondary one which is adapted freely during the continuation to ensure mass conservation.

\section*{Data availability}
The raw data corresponding to the figure shown in this article are available as Supplementary Material \cite{SI}.

\section*{Conflicts of interest}
There are no conflicts of interest to declare.

\acknowledgments{We thank Jens Bickmann, Hauke Hawighorst, and Simon May for helpful discussions. M.t.V. and T.F.H. thank the Studienstiftung des deutschen Volkes for financial support. R.W.\ is funded by the Deutsche Forschungsgemeinschaft (DFG, German Research Foundation) -- 283183152.}

\bibliography{refs}
\end{document}